\documentclass[10pt, conference, letterpaper]{IEEEtran}
\ifCLASSINFOpdf
\else
\fi
\usepackage{graphicx}
\usepackage{algorithmic}
\usepackage{booktabs}
\usepackage{blindtext, graphicx}
\usepackage{amsmath, amsthm}
\usepackage{amsfonts}
\usepackage{stfloats}
\usepackage{subfigure}
\usepackage{slashbox}
\usepackage{algorithm}
\usepackage{mathrsfs}
\usepackage{MnSymbol}
\usepackage{wasysym}
\usepackage{array}
\usepackage{multirow}
\usepackage{threeparttable}
\usepackage{color}

\begin{document}
%
% paper title
% can use linebreaks \\ within to get better formatting as desired
\title{Dynamic Interference Steering in Heterogeneous Cellular Networks}

% author names and affiliations
% use a multiple column layout for up to three different
% affiliations
%\author{\IEEEauthorblockN{Zhao Li$^{*}$, Canyu Shu$^{*}$, Kang G. Shin$^{\dagger}$}
%\IEEEauthorblockA{$^{*}$State Key Laboratory of Integrated Service Networks, Xidian University\\$^{\dagger}$Department of Electrical Engineering and Computer Science, The University of Michigan\\
%zli@xidian.edu.cn, cyshu112@126.com, kgshin@eecs.umich.edu}
%}
\author{\IEEEauthorblockN{Zhao Li$^{*}$, Canyu Shu$^{*}$, Fengjuan Guo$^{*}$, Kang G. Shin$^{\dagger}$, Jia Liu$^{\ddagger}$}
\IEEEauthorblockA{$^{*}$State Key Laboratory of Integrated Service Networks, Xidian University\\
$^{\dagger}$The University of Michigan, USA\\
$^{\ddagger}$National Institute of Informatics, Japan\\
zli@xidian.edu.cn, cyshu112@126.com, fjguo@stu.xidian.edu.cn, kgshin@umich.edu, jliu@nii.ac.jp}}

% conference papers do not typically use \thanks and this command
% is locked out in conference mode. If really needed, such as for
% the acknowledgment of grants, issue a \IEEEoverridecommandlockouts
% after \documentclass

% for over three affiliations, or if they all won't fit within the width
% of the page, use this alternative format:
%
%\author{\IEEEauthorblockN{Michael Shell\IEEEauthorrefmark{1},
%Homer Simpson\IEEEauthorrefmark{2},
%James Kirk\IEEEauthorrefmark{3},
%Montgomery Scott\IEEEauthorrefmark{3} and
%Eldon Tyrell\IEEEauthorrefmark{4}}
%\IEEEauthorblockA{\IEEEauthorrefmark{1}School of Electrical and Computer Engineering\\
%Georgia Institute of Technology,
%Atlanta, Georgia 30332--0250\\ Email: see http://www.michaelshell.org/contact.html}
%\IEEEauthorblockA{\IEEEauthorrefmark{2}Twentieth Century Fox, Springfield, USA\\
%Email: homer@thesimpsons.com}
%\IEEEauthorblockA{\IEEEauthorrefmark{3}Starfleet Academy, San Francisco, California 96678-2391\\
%Telephone: (800) 555--1212, Fax: (888) 555--1212}
%\IEEEauthorblockA{\IEEEauthorrefmark{4}Tyrell Inc., 123 Replicant Street, Los Angeles, California 90210--4321}}

% use for special paper notices
%\IEEEspecialpapernotice{(Invited Paper)}

% make the title area
\maketitle

\begin{abstract}
%\boldmath
With the development of diverse wireless communication technologies,
interference has become a key impediment in network performance,
thus making effective interference management (IM) essential to
accommodate a rapidly increasing number of subscribers with diverse services.
Although there have been numerous IM schemes proposed thus far,
none of them are free of some form of cost.
It is, therefore, important to balance the benefit brought by and cost of each adopted IM scheme by adapting its operating parameters to various network deployments and dynamic channel conditions.

We propose a novel IM scheme, called \textit{dynamic interference steering} (DIS), by recognizing the fact
that interference can be not only suppressed or mitigated but also steered in a particular direction.
Specifically, DIS exploits both channel state information (CSI) and the data contained in the interfering signal
to generate a signal that modifies the spatial feature of the original interference to
partially or fully cancel the interference
appearing at the victim receiver.
By intelligently determining the strength of the steering signal,
DIS can steer the interference in an optimal direction to balance the transmitter's power
used for IS and the desired signal's transmission.
DIS is shown via simulation to be able to make better use of the transmit power,
hence enhancing users' spectral efficiency (SE) effectively.

\end{abstract}

% For peer review papers, you can put extra information on the cover
% page as needed:
% \ifCLASSOPTIONpeerreview
% \begin{center} \bfseries EDICS Category: 3-BBND \end{center}
% \fi
%
% For peerreview papers, this IEEEtran command inserts a page break and
% creates the second title. It will be ignored for other modes.
\IEEEpeerreviewmaketitle

\section{Introduction}
% no \IEEEPARstart
Due to the broadcast nature of wireless communications, concurrent transmissions
from multiple source nodes may cause interferences to each other, thus degrading
subscribers' data rate. Interference management (IM) is, therefore, crucial to
meet the ever-increasing demand of diverse users' Quality-of-Service (QoS).

Interference alignment (IA) is a powerful means of controlling interference
and has thus been under development in recent years.
By preprocessing signals at the transmitter, multiple interfering signals are mapped into
a certain signal subspace, i.e., the overall interference space at the receiver is minimized,
leaving the remaining subspace interference-free [1-2].
IA is shown to be able to achieve the information-theoretic maximum DoF
(Degree of Freedom) in some interference networks [3-4].
However, to achieve such a promising gain,
it is required to use either infinite symbol extensions over time/frequency [4]
or a large number of antennas at each receiver [5], both which are not realistic.
That is, IA emerges as a promising IM scheme, but its applicability is severely limited
by the high DoF requirement.

Some researchers have attempted to circumvent the stringent DoF requirement
by proposing other IM schemes, such as interference neutralization (IN).
IN refers to the distributed zero-forcing of interference
when the interfering signal traverses multiple nodes
before arriving at the undesired receivers/destinations.
The basic idea of IN has been applied to deterministic channels %\textcolor{blue}{[6-7]}
and interference networks [6-10], both of which employ relays.
IM was studied in the context of a deterministic wireless interaction model [6-7].
In [6], IN was proposed and an admissible rate region of
the Gaussian ZS and ZZ interference-relay networks was obtained, where
ZS and ZZ denote two special configurations of a two-stage interference-relay
network in which some of the cross-links are weak.
The authors of [7] further translated the exact capacity region obtained in [6]
into a universal characterization for the Gaussian network.
The key idea used in the above interference networks with relays is
to control the precoder at the relay so that the sum of the channel gains
of the newly-created signal path via the relay and
the direct path to the destination becomes zero.
A new scheme called {\em aligned interference neutralization} was proposed in [8]
by combining IA and IN. It provides a way to align interference terms
over each hop so as to cancel them over the air at the last hop.
However, this conventional relay incurs a processing delay --- while the direct path
does not --- between a source-destination pair, limiting the DoF gain in a wireless
interference network. To remedy this problem, an instantaneous relay (or relay-without-delay)
was introduced in [9-10] to obtain a larger capacity than the conventional relay,
and a higher DoF gain was achieved without requiring any memory at the relay.
Although the studies based on this instantaneous relay provide some useful theoretical results,
this type of relay is not practical.
On the other hand, IN can mitigate interference, but the power overhead of
generating neutralization signals also affects the system's performance.
To the best of our knowledge, the power overhead has not been considered in any of the existing
studies related to IN --- either more recent interference neutralization [8-10] or
the same ideas known via other names for many years, such as distributed orthogonalization,
distributed zero-forcing, multiuser zero-forcing, and orthogonalize-and-forward [11-12].
In practice, higher transmit power will be used by IN when the interference is strong,
thus making less power available for the desired data transmission.
Furthermore, IN may not even be available for mobile terminals due to their limited power budget.

By recognizing the fact that interference can be not only neutralized but also steered in a particular direction,
one can ``steer'' interference, which we call {\em interference steering} (IS) [13].
That is, a steering signal is generated to modify the interference's spatial feature
so as to steer the original interference in the direction orthogonal to that
of the desired signal perceived at the victim receiver.
Note, however, that IS simply steers the original interference in the direction orthogonal
to the desired signal, which we call {\em Orthogonal-IS} (OIS) in the
following discussion, regardless of the underlying channel conditions,
as well as the strength and spatial feature of interference(s) and the intended transmission.
%=======================================
% 2016.7.19 ORIGINAL DESCRIPTION
%By recognizing that interference can be not only neutralized but also steered in a particular direction,
%we proposed a new IM technique called {\em interference steering} (IS) and applied it to an
%infrastructure-based enterprise wireless local area network (WLAN) [13].
%With IS, a steering signal is generated to modify the interference's spatial feature
%so that the original interference is steered in the orthogonal direction
%of the desired signal observed at the victim.
%Compared to IN, IS consumes much less transmit power [13],
%yet a spatial DoF is required to place the steered interference
%which is similar to the principle of IA.
%However, in the proposed IS scheme, the original interference is simply steered in
%the direction orthogonal the desired signal, which we call {\em Orthogonal-IS} (OIS) in the
%following discussion, regardless of the various channel conditions
%as well as the strength and spatial feature of interference(s) and the intended transmission.}
% 2016.7.19 ORIGINAL DESCRIPTION
%=======================================
Therefore, the tradeoff between the benefit of IS (i.e., interference suppression) and its power cost
was not considered there.
Since the more transmit power is spent on interference steering,
the less power for the desired signal's transmission will be available,
one can naturally raise a question:
``Is it always necessary to steer interference in the direction orthogonal to the desired signal?''

To answer the above question, we propose a new IM scheme, called \textit{dynamic interference steering} (DIS).
With DIS, the spatial feature of the steered interference at the intended receiver
is intelligently determined so as to achieve a balance between the transmit power consumed by IS and the
residual interference due to the imperfect interference suppression, thus improving the user's SE.

The contributions of this paper are two-fold:

\begin{itemize}
\item Proposal of a novel IM scheme called \textit{dynamic interference steering} (DIS).
By intelligently determining the strength of steering signal, we balance the transmit power
used for IS and that for the desired signal's transmission. DIS can also subsume orthogonal-IS
as a special case, making it more general.
\item Extension of DIS to general cases where the number of interferences from macro base station (MBS),
the number of desired signals from a pico base station (PBS) to its intended pico user equipment (PUE),
and the number of PBSs and PUEs are all variable.
\end{itemize}

The rest of this paper is organized as follows.
Section II describes the system model, while Section III details the dynamic interference steering.
Section IV presents the generalization of DIS and
Section V evaluates its performance and overhead.
Finally, Section VI concludes the paper.

Throughout this paper, we use the following notations. The set of complex numbers
is denoted as $\mathbb{C}$, while vectors and matrices are represented by bold
lower-case and upper-case letters, respectively. Let $\mathbf{X}^{T}$,
$\mathbf{X}^{H}$ and $\mathbf{X}^{-1}$ denote the transpose, Hermitian,
and inverse of matrix $\mathbf{X}$.
$\|\cdot\|$ and $|\cdot|$ indicate the Euclidean norm and the absolute value.
$\mathbb{E}(\cdot)$ denotes statistical expectation and $\langle\mathbf{a}, \mathbf{b}\rangle$
represents the inner product of two vectors.

\section{System Model}

We consider the downlink\footnote{Neither IS nor DIS is applicable for uplink transmission due to
their requirement of Tx's cooperation.} transmission in heterogeneous cellular networks (HCNs)
composed of overlapping macro and pico cells [14].
As shown in Fig.~1, macro and pico base stations (MBSs and PBSs)
are equipped with $N_{T_{1}}$ and $N_{T_{0}}$ antennas,
whereas macro user equipment (MUE) and PUE are equipped with
$N_{R_{1}}$ and $N_{R_{0}}>1$ antennas, respectively.
Since mobile stations/devices are subject to severer restrictions in cost and hardware,
than a base station (BS), the BS is assumed to have no less antennas than a UE,
i.e., $N_{T_{i}}\geq N_{R_{i}}$ where $i=0,1$.
The radio range, $d$, of a picocell is known to be 300$m$ or less,
whereas the radius, $D$, of a macrocell is around 3000$m$ [14].
Let $\mathbf{x}_{1}$ and $\mathbf{x}_{0}$ denote the transmit data vectors
from MBS and PBS to their serving subscribers, respectively.
$\mathbb{E}(\|\mathbf{x}_{1}\|^{2})=\mathbb{E}(\|\mathbf{x}_{0}\|^{2})=1$ holds.
For clarity of exposition, our design begins with the assumption of
beamforming (BF), i.e., only one data stream
is sent from MBS to MUE (or from PBS to PUE).
Then, $\mathbf{x}_{1}$ and $\mathbf{x}_{0}$ become scalars $x_{1}$ and $x_{0}$.
We will generalize this to multiple data streams sent from MBS and PBS in Section IV.
We use $P_{1}$ and $P_{0}$ to denote the transmit power of MBS and PBS, respectively.
Let $\mathbf{H}_{0}\in\mathbb{C}^{N_{R_{0}}\times N_{T_{0}}}$ and
$\mathbf{H}_{1}\in\mathbb{C}^{N_{R_{1}}\times N_{T_{1}}}$ be the channel matrices
from MBS to MUE and from PBS to PUE, respectively, whereas that from MBS to PUE
is denoted by $\mathbf{H}_{10}\in\mathbb{C}^{N_{R_{0}}\times N_{T_{1}}}$.
We adopt a spatially uncorrelated Rayleigh flat fading channel model
to model the elements of the above matrices
as independent and identically distributed zero-mean unit-variance
complex Gaussian random variables.
We assume that all users experience block fading,
i.e., channel parameters remain constant in a block consisting of
several successive time slots and vary randomly between successive blocks.
Each user can accurately estimate CSI
w.r.t. its intended and unintended Txs
and feed it back to the associated BS via a low-rate
error-free link.
We assume reliable links for the delivery of CSI and signaling.
The delivery delay is negligible relative to the time scale
on which the channel state varies.
\begin{figure}[!htb]
\graphicspath{/fig}
\centering
\vspace*{-0.1in}
\includegraphics[width=0.43\textwidth]{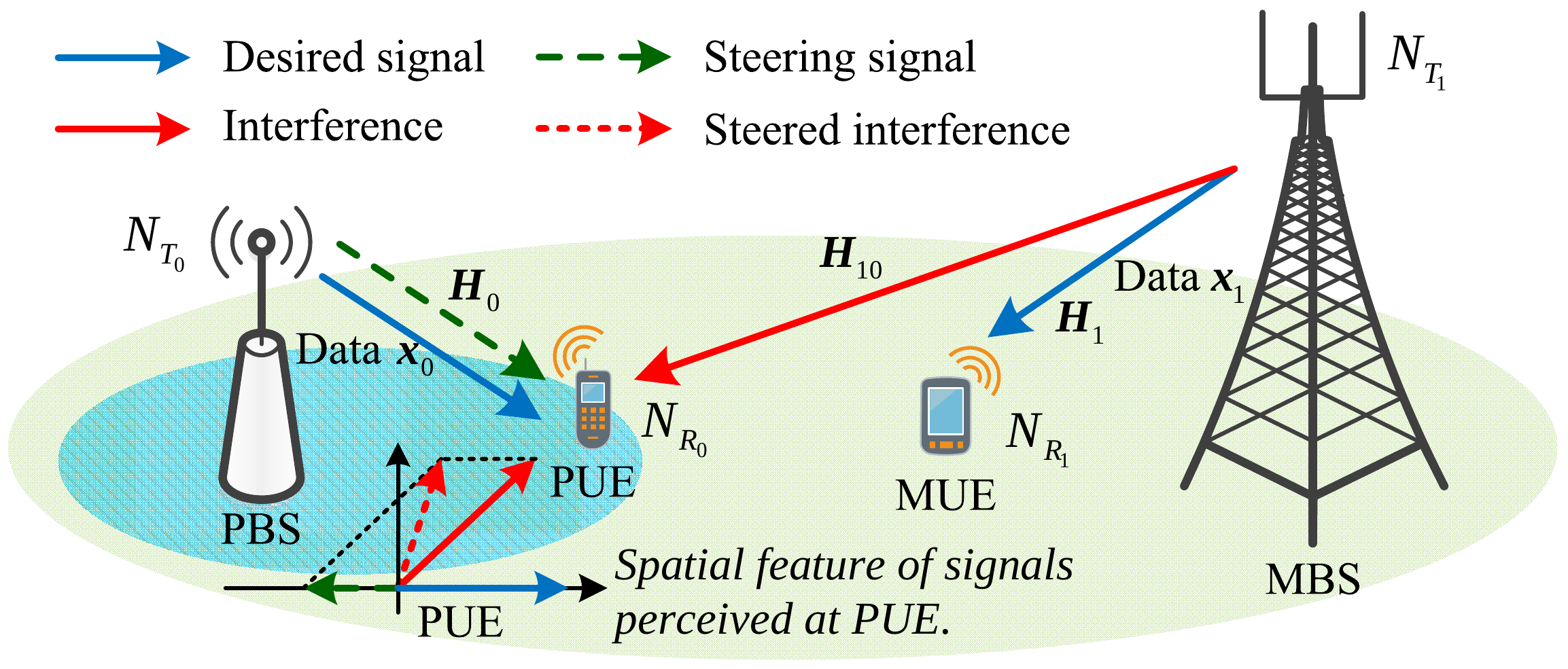}
\vspace{-5pt}
\caption{System model.}
\label{fig: figure one}
\vspace*{-0.1in}
\end{figure}

As mobile data traffic has increased significantly in recent years,
network operators have strong preference of open access to
offload users' traffic from heavily loaded macrocells to
other infrastructures such as picocells [14-15].
Following this trend, we assume each PBS operates in an open mode,
i.e., users in the coverage of a PBS are allowed to access it.
The transmission from MBS to MUE will interfere with
the intended transmission from PBS at PUE.
Nevertheless, due to the limited coverage of a picocell,
PBS will not cause too much interference to MUE,
and is thus omitted in this paper. As a result,
the interference shown in Fig.~1 is asymmetric.

Since picocells are deployed to improve the capacity and
coverage of existing cellular systems,
each picocell has subordinate features as compared to the macrocell,
and hence the macrocell transmission is given priority over the picocell's transmission.
Specifically, MBS will not adjust its transmission for pico-users.
However,
we assume that PBS can acquire the information of $x_{1}$ via inter-BS collaboration;
this is easy to achieve because PBS and MBS are deployed by the same operator [16].
With such information, DIS can be implemented to adjust the disturbance in a proper direction at PUE.
Since the transmission from MBS to MUE depends only on $\mathbf{H}_{1}$ and
is free from interference, we only focus on the pico-users' transmission performance.

Although we take HCN as an example to design our scheme,
it should be noticed that other types of network as long as they are featured as
1) collaboration between the interfering Tx and victim Tx is available,
and 2) the interference topology is asymmetric, our scheme is applicable.

\section{Dynamic Interference Steering}

As mentioned earlier, by generating a duplicate of the interference
and sending it along with the desired signal,
the interference could be steered in the direction orthogonal
to the desired signal at the intended PUE with orthogonal-IS.
However, the tradeoff between the benefit of interference steering
and its power cost has not been considered before.
That is, under a transmit power constraint, the more power consumed for IS,
the less power will be available for the intended signal's transmission.
To remedy this deficiency, we propose a novel IM scheme called \textit{dynamic interference steering} (DIS).
By intelligently determining the strength of steering signal,
the original interference is adjusted in an appropriate direction.
DIS balances the transmit power used for generating the steering signal
and that for the desired signal's transmission.

\subsection{Signal Processing of DIS}

As mentioned above, since the macrocell receives higher priority than picocells,
MBS will not adjust its transmission for pico-users.
In what follows,
we use $N_{T_{i}}=N_{R_{i}}\geq 2$ where $i=0,1$ as an example,
but our scheme can be easily extended to the case of $N_{T_{i}}\geq N_{R_{i}}$.

Due to path loss, the mixed signal received at PUE can be expressed as:
\begin{equation}
\mathbf{r}_{0}=\sqrt{P_{0}10^{-0.1L_{0}}}\mathbf{H}_{0}\mathbf{p}_{0}{x}_{0}
+\sqrt{P_{1}10^{-0.1L_{10}}}\mathbf{H}_{10}\mathbf{p}_{1}x_{1}
+\mathbf{n}_{0}
%\vspace*{-0.05in}
\end{equation}
where the column vectors $\mathbf{p}_{0}$ and $\mathbf{p}_{1}$
represent the precoders for data symbols $x_{0}$ and $x_{1}$
sent from PBS and MBS, respectively.
The first term on the right hand side (RHS) of Eq.~(1) is the desired signal,
the second term denotes the interference from MBS, and $\mathbf{n}_{0}$
represents for the additive white Gaussian noise (AWGN) with zero-mean and variance $\sigma^{2}_{n}$.
The path loss from MBS and PBS to a PUE %a mobile terminal
is modeled as
%$L_{(\cdot)}=128.1+37.6log_{10}[\eta_{(\cdot)}/10^{3}]~dB$ and
%$L_{(\cdot)}=38+30log_{10}[\eta_{(\cdot)}]~dB$,
$L_{10}=128.1+37.6\log_{10}(\eta_{10}/10^{3})~dB$ and
$L_{0}=38+30\log_{10}(\eta_{0})~dB$,
respectively [17],
where the variable $\eta_{(\cdot)}$, measured in meters ($m$), is the distance from the transmitter to the receiver.

The estimated signal at PUE after post-processing can be written as
$\tilde{r}_{0}=\mathbf{f}^{H}_{0}\mathbf{r}_{0}$ where $\mathbf{f}_{0}$ denotes the receive filter.
Recall that the picocell operates in an open mode, and a MUE in the area covered
by PBS will become a PUE and then be served by the PBS.
The interference model shown in Fig.~1 has an asymmetric feature
in which only the interference from MBS to PUE is considered.
Moreover, since the macrocell is given priority over the picocells,
MBS will not adjust its transmission for the PUEs, and hence
transmit packets to MUE based only on $\mathbf{H}_{1}$.
Here we adopt the singular value decomposition (SVD) based BF transmission,
but can also use other types of pre- and post-processing.
Applying SVD to $\mathbf{H}_{i}$ ($i=0,1$), we get
$\mathbf{H}_{i}=\mathbf{U}_{i}\mathbf{\Sigma}_{i}\mathbf{V}^{H}_{i}$.
We then employ $\mathbf{p}_{i}=\mathbf{v}^{(1)}_{i}$ and $\mathbf{f}_{i}=\mathbf{u}^{(1)}_{i}$,
where $\mathbf{v}^{(1)}_{i}$ and $\mathbf{u}^{(1)}_{i}$ are the first column vectors
of the right and left singular matrices ($\mathbf{V}_{i}$ and $\mathbf{U}_{i}$), respectively,
both of which correspond to the principal eigen-mode of $\mathbf{H}_{i}$.

From Fig.~1 one can see that the strengths of desired signal and interference at PUE depend on
the network topology, differences of transmit power at PBS and MBS, as well as channel conditions.
All of these factors affect the effectiveness of IM.
For clarity of presentation, we define $P^{e}_{0}=P_{0}10^{-0.1L_{0}}$, $P^{e}_{1}=P_{1}10^{-0.1L_{10}}$,
where $P^{e}_{0}$ and $P^{e}_{1}$ indicate the transmit power of PBS and MBS
incorporated with the path loss perceived by PUE. With this definition,
consideration of various network topologies and transmit power differences
can be simplified to $P^{e}_{0}$ and $P^{e}_{1}$. In what follows,
we first present the basic principle of orthogonal-IS (OIS),
and then elaborate on the design of dynamic-IS (DIS) where we provide
the existence and calculation of the optimal steering signal.
\begin{figure}[!htb]
\vspace*{-0.1in}
\graphicspath{/fig}
\centering
\subfigure[An illustration of OIS.]{
\includegraphics[width=1.25in,height=0.95in]{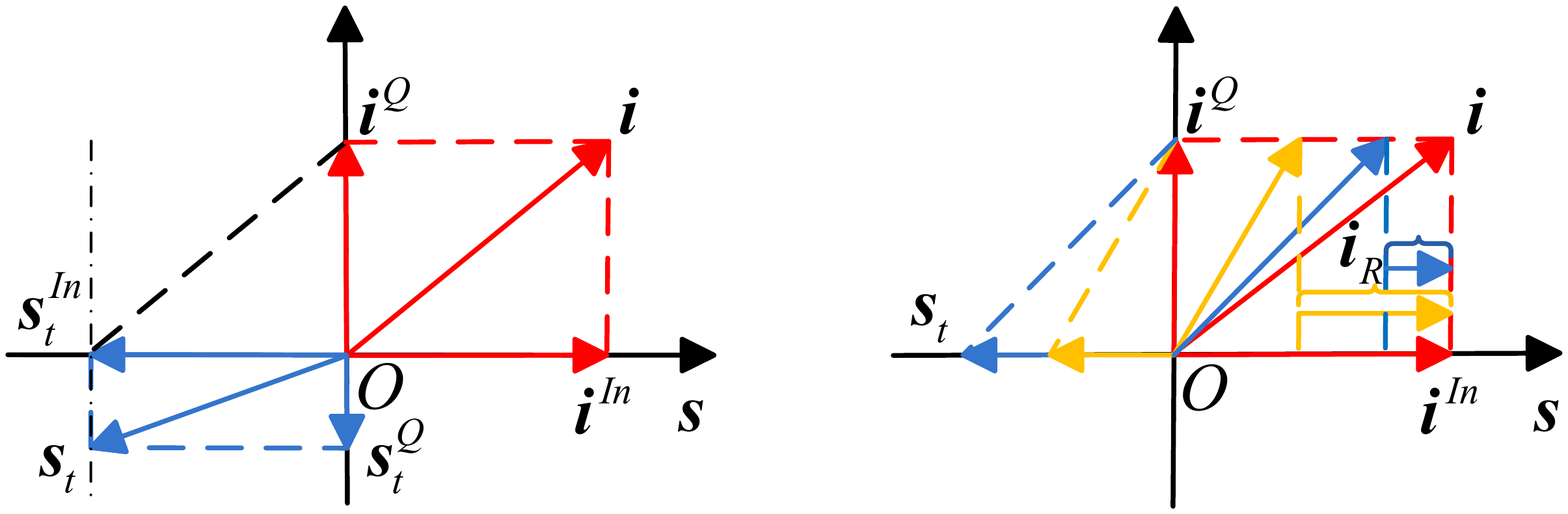}}
\hspace{0.2in}
\subfigure[An illustration of DIS.]{
\includegraphics[width=1.25in,height=0.95in]{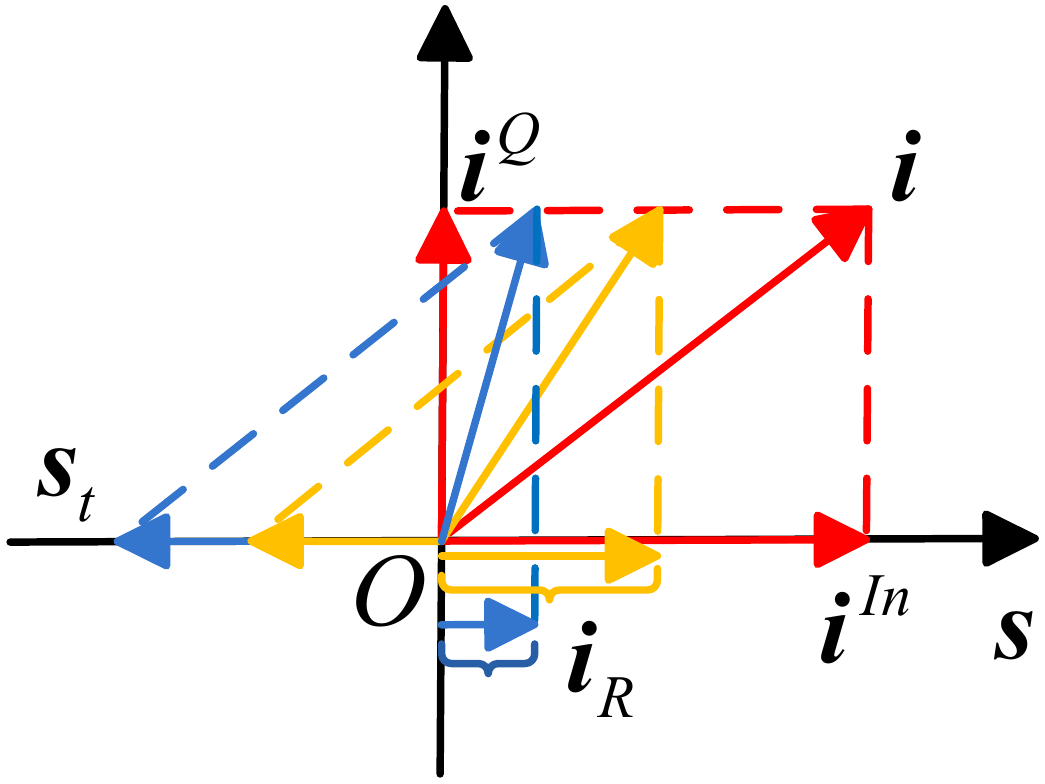}}
\vspace{-3pt}
\caption{Illustrations of OIS and DIS.}
\label{fig: figure two} %% label for entire figure
\vspace*{-0.07in}
\end{figure}

With OIS, PBS acquires interference information, including data and CSI,
from MBS via inter-BS collaboration and by PUE's estimation and feedback, respectively.
PBS then generates a duplicate of the interference and sends it along with the desired signal.
The former is used for interference steering at PUE,
whereas the latter carries the payload.
The received signal at PUE then becomes:
\begin{equation}
%\vspace*{-0.01in}
\begin{aligned}
\mathbf{r}_{0}&=\sqrt{P^{e}_{0}-P^{e}_{OIS}}\mathbf{H}_{0}\mathbf{p}_{0}{x}_{0}
+\sqrt{P^{e}_{1}}\mathbf{H}_{10}\mathbf{p}_{1}x_{1}\\
&+\sqrt{P^{e}_{OIS}}\mathbf{H}_{0}\mathbf{p}_{OIS}x_{1}+\mathbf{n}_{0}
\end{aligned}
%\vspace*{-0.01in}
\end{equation}
where $P^{e}_{OIS}=P_{OIS}10^{-0.1L_{0}}$.
$P_{OIS}$ represents the power overhead of OIS,
and $\mathbf{p}_{OIS}$ is the precoder for the steering signal.
We first define the directions of the desired signal and
the original interference combined with the steering signal as
$\mathbf{d}_{\mathbf{s}}=\frac{\mathbf{H}_{0}\mathbf{p}_{0}}{\|\mathbf{H}_{0}\mathbf{p}_{0}\|}$ and
$\mathbf{d}_{\mathbf{i}+\mathbf{s}_{t}}=\frac{\sqrt{P^{e}_{1}}\mathbf{H}_{10}\mathbf{p}_{1}+\sqrt{P^{e}_{OIS}}
\mathbf{H}_{0}\mathbf{p}_{OIS}}{\|\sqrt{P^{e}_{1}}\mathbf{H}_{10}\mathbf{p}_{1}+\sqrt{P^{e}_{OIS}}
\mathbf{H}_{0}\mathbf{p}_{OIS}\|}$, respectively.
Then, the original interference should be steered in the direction orthogonal to the desired signal
by letting $\langle\mathbf{d}_{\mathbf{s}},\mathbf{d}_{\mathbf{i}+\mathbf{s}_{t}}\rangle=0$.
As shown in Fig.~2(a), both the disturbance ($\mathbf{i}$) and
the steering signal ($\mathbf{s}_{t}$),
can be decomposed into an in-phase component and a quadrature component,
denoted by the superscripts $In$ and $Q$, respectively,
w.r.t. the intended transmission $\mathbf{s}$,
i.e., $\mathbf{i}=\mathbf{i}^{In}+\mathbf{i}^{Q}$ and $\mathbf{s}_{t}=\mathbf{s}^{In}_{t}+\mathbf{s}^{Q}_{t}$.
When $\mathbf{s}^{In}_{t}=-\mathbf{i}^{In}$, OIS is realized.
Furthermore, since the length of a vector indicates the signal's strength, OIS with the minimum power
overhead is achieved when $\mathbf{s}_{t}=\mathbf{s}^{In}_{t}$, i.e., $\mathbf{s}^{Q}_{t}=\mathbf{0}$.
Hence, in order to reduce power cost, we let $\mathbf{s}^{Q}_{t}=\mathbf{0}$.
It can be easily seen that $\mathbf{i}^{In}=\sqrt{P^{e}_{1}}\mathbf{P}\mathbf{H}_{10}\mathbf{p}_{1}$
where $\mathbf{P}=\mathbf{d}_{\mathbf{s}}(\mathbf{d}^{T}_{\mathbf{s}}\mathbf{d}_{\mathbf{s}})^{-1}
\mathbf{d}^{T}_{\mathbf{s}}$ denotes the projection matrix.
To implement OIS, the steering signal should satisfy
$\sqrt{P^{e}_{OIS}}\mathbf{H}_{0}\mathbf{p}_{OIS}=-\sqrt{P^{e}_{1}}\mathbf{P}\mathbf{H}_{10}\mathbf{p}_{1}$.
This equation can be decomposed into
$\mathbf{H}_{0}\mathbf{p}_{OIS}=-\alpha\mathbf{P}\mathbf{H}_{10}\mathbf{p}_{1}$ and
$P^{e}_{OIS}=\beta P^{e}_{1}$ where $\alpha\beta=1$,
%From $\mathbf{H}_{0}\mathbf{p}_{OIS}=-\alpha\mathbf{P}\mathbf{H}_{10}\mathbf{p}_{1}$,
so that we can get $\mathbf{p}_{OIS}=-\alpha\mathbf{H}^{-1}_{0}\mathbf{P}\mathbf{H}_{10}\mathbf{p}_{1}$.
Note that $\|\mathbf{p}_{OIS}\|=1$ is not guaranteed,
i.e., $\mathbf{p}_{OIS}$ affects the power cost of OIS.
%==========================
% 2017.1.30 Bakcup
%This equation can be decomposed into two parts,
%$\mathbf{H}_{0}\mathbf{p}_{OIS}=-\mathbf{P}\mathbf{H}_{10}\mathbf{p}_{1}$ and
%$P^{e}_{OIS}=P^{e}_{1}\|\mathbf{p}_{OIS}\|^2$,
%where $\mathbf{p}_{OIS}=-\mathbf{H}^{-1}_{0}\mathbf{P}\mathbf{H}_{10}\mathbf{p}_{1}$.
%Note that $\|\mathbf{p}_{OIS}\|=1$ is not guaranteed,
%i.e., $\mathbf{p}_{OIS}$ has impact on the power cost of OIS.
%==========================

When $N_{T_{i}}>N_{R_{i}}$ ($i=0,1$), the inverse of $\mathbf{H}_{0}$ should be
replaced by its Moore-Penrose pseudo-inverse. The mechanism can then be generalized.
In addition, when the interference is too strong, $P_0$ may not be sufficient for OIS,
in such a case, we can simply switch to the non-interference management (non-IM) mode,
e.g., matched filtering (MF) at the victim receiver,
or other IM schemes with less or no transmit power consumption such as zero-forcing reception.

By adopting $\mathbf{f}_{0}=\mathbf{u}^{(1)}_{0}$ as the receive filter,
the SE of PUE with OIS can be computed as:
\begin{equation}
\vspace*{-0.05in}
\setcounter{equation}{3}
c^{OIS}_{0}=\log_{2}\left\{
1+\frac{(P^{e}_{0}-P^{e}_{OIS})[\lambda^{(1)}_{0}]^{2}}
{\sigma^{2}_{n}}
\right\},
%\vspace*{-0.05in}
\end{equation}
where $\lambda^{(1)}_{0}$ is the largest singular value of $\mathbf{H}_{0}$,
indicating the amplitude gain of the principal spatial sub-channel.

From Eq.~(3), we can see that although the original interference is
steered into the orthogonal direction of the desired signal and
the disturbance to the intended transmission is completely eliminated,
it accompanies a transmit power loss, $P^{e}_{OIS}$,
degrading the received desired signal strength.
One can then raise a question: ``is the orthogonal-IS always necessary/worthwhile?''
To answer this question, we may adjust both the direction and
strength of the steering signal adaptively to implement dynamic IS.
Note, however, that in order to minimize the transmit power overhead,
the steering signal should be opposite to the spatial feature of the desired signal.
Thus, only the strength of steering signal needs to be adjusted.
In what follows, we generalize the OIS to DIS by
introducing a coefficient $\rho\in (0,1]$ called the \textit{steering factor},
representing the portion of in-phase component of the disturbance
w.r.t. the desired signal to be mitigated.
When $\rho=1$, orthogonal-IS is realized,
while DIS approaches non-IM as $\rho\rightarrow 0$.
Without ambiguity, we adopt $\mathbf{s}_{t}$ to represent the dynamic steering signal.
Then, we have
$\mathbf{s}_{t}=-\rho\mathbf{i}^{In}=-\rho\sqrt{P^{e}_{1}}\mathbf{P}\mathbf{H}_{10}\mathbf{p}_{1}$.
As illustrated by Fig.~2(b), when $\rho<1$, interference $\mathbf{i}$ is steered into a direction
not orthogonal to $\mathbf{d}_{\mathbf{s}}$,
i.e., $\mathbf{i}^{In}$ is not completed eliminated.
Then, the steered interference becomes $\mathbf{i}+\mathbf{s}_{t}=(1-\rho)\mathbf{i}^{In}+\mathbf{i}^{Q}$,
whose projection on $\mathbf{d}_{\mathbf{s}}$ is non-zero,
i.e., provided that $\rho<1$, a residual interference,
expressed as $\mathbf{i}_{R}=\mathbf{i}^{In}+\mathbf{s}_{t}=(1-\rho)\mathbf{i}^{In}$, exists.

Similarly to the discussion about OIS, to implement DIS, the following equation should hold:
\begin{equation}
\vspace{-0.05in}
\sqrt{P^{e}_{DIS}}\mathbf{H}_{0}\mathbf{p}_{DIS}=-\rho\sqrt{P^{e}_{1}}\mathbf{P}\mathbf{H}_{10}\mathbf{p}_{1}.
%\vspace{-0.05in}
\end{equation}

For clarity of exposition, we normalize the precoder so that the direction and strength requirements for DIS
could be decoupled from each other. Then, the expression of DIS implementation is given as:
\begin{equation}
\vspace*{-0.05in}
\left\{
\begin{array}{l}
\mathbf{p}_{DIS}=-\mathbf{H}^{-1}_{0}\mathbf{P}\mathbf{H}_{10}\mathbf{p}_{1}/
\|\mathbf{H}^{-1}_{0}\mathbf{P}\mathbf{H}_{10}\mathbf{p}_{1}\|\\
P^{e}_{DIS}=\rho^{2}P^{e}_{1}\|\mathbf{H}^{-1}_{0}\mathbf{P}\mathbf{H}_{10}\mathbf{p}_{1}\|^2
\end{array}
\right.,
%\vspace*{-0.05in}
\end{equation}
where $\mathbf{p}_{DIS}$ is the precoder for steering signal and
$P^{e}_{DIS}$ denotes the power overhead for DIS at PBS, i.e., $P_{DIS}$,
incorporated with path loss $10^{-0.1L_{0}}$.

The received signal at PUE with DIS is then
\begin{equation}
%\vspace*{-0.02in}
\mathbf{r}_{0}=\sqrt{P^{e}_{0}-P^{e}_{DIS}}\mathbf{H}_{0}\mathbf{p}_{0}{x}_{0}
+\sqrt{P^{e}_{1}}(1-\rho)\mathbf{P}\mathbf{H}_{10}\mathbf{p}_{1}x_{1}+\mathbf{n}_{0},
%\vspace*{-0.02in}
\end{equation}
where the second term on the RHS of Eq.~(6) indicates the residual interference,
which is the in-phase component of the original interference
$\sqrt{P^{e}_{1}}\mathbf{H}_{10}\mathbf{p}_{1}x_{1}$ combined with the steering signal
$\rho\sqrt{P^{e}_{DIS}}\mathbf{H}_{0}\mathbf{p}_{DIS}x_{1}$
w.r.t. the desired transmission.

By employing $\mathbf{f}_{0}=\mathbf{u}^{(1)}_{0}$ as the receive filter,
the achievable SE of PUE employing DIS can then be calculated as:
\begin{equation}
\vspace*{-0.05in}
c^{DIS}_{0}=\log_{2}\left\{
1+\frac{(P^{e}_{0}-P^{e}_{DIS})[\lambda^{(1)}_{0}]^{2}}
{\sigma^{2}_{n}+I_{R}}
\right\},
%\vspace*{-0.06in}
\end{equation}
where $I_{R}=P^{e}_{1}\|\mathbf{f}^{H}_{0}(1-\rho)\mathbf{P}\mathbf{H}_{10}\mathbf{p}_{1}\|^2$
denotes the strength of residual interference after post-processing at PUE.

Based on the above discussion, it can be easily seen that $I_{R}=0$ when $\rho=1$,
i.e., DIS becomes OIS. So, DIS includes OIS as a special case, making it more general.

\subsection{Optimization of Steering Factor $\rho$}

In what follows, we will discuss the existence of the optimal $\rho$,
denoted by $\rho^{*}$, with which PUE's SE can be maximized with limited $P_{0}$.
Based on the Shannon's equation, we can instead optimize the signal-to-interference-plus-noise
ratio (SINR) of PUE, denoted by $\varphi_{0}$.

Substituting Eq.~(5) into Eq.~(7), we can obtain $\varphi_{0}$ as:
\begin{equation}
\begin{aligned}
\varphi_{0}&=\frac{\left(P^{e}_{0}-\rho^2P^{e}_{1}\|\mathbf{H}^{-1}_{0}\mathbf{P}\mathbf{H}_{10}
\mathbf{p}_{1}\|^2\right)[\lambda^{(1)}_{0}]^2}
{(1-\rho)^2P^{e}_{1}\|\mathbf{f}^{H}_{0}\mathbf{P}\mathbf{H}_{10}\mathbf{p}_{1}\|^2+\sigma^2_{n}}\\
&=\frac{P^{e}_{0}[\lambda^{(1)}_{0}]^2-\rho^2P^{e}_{1}\|\mathbf{g}\|^2[\lambda^{(1)}_{0}]^2}{\rho^2P^{e}_{1}
|\chi|^2-(2\rho-1)P^{e}_{1}|\chi|^2+\sigma^2_{n}}
\end{aligned},
\end{equation}
where $\mathbf{g}=\mathbf{H}^{-1}_{0}\mathbf{P}\mathbf{H}_{10}\mathbf{p}_{1}$ and
$\chi=\mathbf{f}^{H}_{0}\mathbf{P}\mathbf{H}_{10}\mathbf{p}_{1}$.

Eq.~(8) can be simplified as:
\begin{equation}
\vspace*{-0.05in}
\varphi_{0}=\frac{A-\rho^2 B}{C-\rho D+\rho^2 E}
%\vspace*{-0.05in}
\end{equation}
where $A=P^{e}_{0}[\lambda^{(1)}_{0}]^{2}$,
$B=P^{e}_{1}\|\mathbf{g}\|^{2}[\lambda^{(1)}_{0}]^{2}$,
$C=P^{e}_{1}|\chi|^{2}+\sigma^{2}_{n}$,
$D=2P^{e}_{1}|\chi|^{2}$ and $E=P^{e}_{1}|\chi|^{2}$.
Note that all of these coefficients are positive.
\begin{figure*}[hb]
\vspace*{-0.05in}
\setcounter{equation}{11}
\begin{equation}
\begin{aligned}
BC+AE-\sqrt{AB}D&=P^{e}_{1}\|\mathbf{g}\|^{2}[\lambda^{(1)}_{0}]^{2}(\sigma^{2}_{0}+P^{e}_{1}|\chi|^{2})
+P^{e}_{0}[\lambda^{(1)}_{0}]^{2}P^{e}_{1}|\chi|^{2}-2\sqrt{P^{e}_{0}[\lambda^{(1)}_{0}]^{2}P^{e}_{1}
\|\mathbf{g}\|^{2}[\lambda^{(1)}_{0}]^{2}}P^{e}_{1}|\chi|^{2}\\
&>(P^{e}_{1})^{2}\|\mathbf{g}\|^{2}[\lambda^{(1)}_{0}]^{2}|\chi|^{2}+P^{e}_{0}P^{e}_{1}[\lambda^{(1)}_{0}]^{2}
|\chi|^{2}-2(P^{e}_{0})^{\frac{1}{2}}(P^{e}_{1})^{\frac{3}{2}}[\lambda^{(1)}_{0}]^{2}\|\mathbf{g}\||\chi|^{2}\\
&=P^{e}_{1}[\lambda^{(1)}_{0}]^{2}|\chi|^{2}\left[(P^{e}_{1})^{\frac{1}{2}}\|\mathbf{g}\|-(P^{e}_{0})^{\frac{1}{2}}\right]^{2}\geq 0
\end{aligned}.
\end{equation}
\vspace*{-0.05in}
\end{figure*}

Next, we elaborate on the existence of $\rho^{*}$ under the $P_{0}$ constraint,
with which $\varphi_{0}$ is maximized.
By substituting $\mathbf{g}$ into Eq.~(5), we can see that
when $P^{e}_{0}>P^{e}_{1}\|\mathbf{g}\|^{2}$,
PBS has enough power to steer the interference into the direction orthogonal to the desired signal,
i.e., OIS is achievable.
Otherwise, when $P^{e}_{0}\leq P^{e}_{1}\|\mathbf{g}\|^{2}$,
the maximum of $\rho$, denoted by $\rho_{max}$ is limited by the PBS's transmit power.
In such a case, $\sqrt{\frac{P^{e}_{0}}{P^{e}_{1}\|\mathbf{g}\|^2}}$ is
the maximum portion of the in-phase component of the original interference w.r.t. the desired signal
that can be mitigated with $P_{0}$.
Based on the above discussion, we have $\rho^{*}\in (0,\rho_{max}]$ where
$\rho_{max}=\min\left(1,\sqrt{\frac{P^{e}_{0}}{P^{e}_{1}\|\mathbf{g}\|^2}}\right)$.
In what follows, we first prove the solvability of $\rho^{*}$,
and then show the quality of the resulting solution(s).

By computing the derivative of $\varphi_{0}$ to $\rho$ and setting it to $0$, we get:
\setcounter{equation}{9}
\begin{equation}
\frac{\frac{BD}{2}\rho^2-(BC+AE)\rho+\frac{AD}{2}}{\frac{1}{2}(C-D\rho+\rho^2E)^2}=0.
\end{equation}

Since the denominator cannot be $0$, we only need to solve
\begin{equation}
\frac{BD}{2}\rho^2-(BC+AE)\rho+\frac{AD}{2}=0
\end{equation}
which is a quadratic equation with one unknown.
Let's define $\Delta=(BC+AE)^{2}-ABD^{2}$.
Since $\Delta=(BC+AE+\sqrt{AB}D)(BC+AE-\sqrt{AB}D)$,
and $BC+AE+\sqrt{AB}D$ is positive, we only need to show that
$BC+AE-\sqrt{AB}D>0$. The proof is given below by Eq.~(12).

Based on the above discussion,
we can obtain two solutions $\rho_{\pm}^{*}=\frac{(BC+AE)\pm\sqrt{\Delta}}{BD}$.
We then need to verify the qualification of the two solutions $\rho_{\pm}^{*}$.
We first investigate the feasibility of the larger solution
$\rho_{+}^{*}=\frac{(BC+AE)+\sqrt{\Delta}}{BD}$.
The first term of $\rho_{+}^{*}$ can be rewritten as:
\setcounter{equation}{12}
\begin{equation}
\begin{aligned}
\frac{BC+AE}{BD}&>\frac{P^{e}_{1}\|\mathbf{g}\|^{2}[\lambda^{(1)}_{0}]^{2}P^{e}_{1}|\chi|^{2}+P^{e}_{0}
[\lambda^{(1)}_{0}]^{2}P^{e}_{1}|\chi|^{2}}
{2P^{e}_{1}\|\mathbf{g}\|^{2}[\lambda^{(1)}_{0}]^{2}P^{e}_{1}|\chi|^{2}}\\
&=\frac{1}{2}+\frac{P^{e}_{0}}{2P^{e}_{1}\|\mathbf{g}\|^{2}}
\end{aligned}.
%\vspace*{-0.05in}
\end{equation}

The second term of $\rho_{+}^{*}$ is:
\begin{equation}
\begin{aligned}
\frac{\sqrt{\Delta}}{BD}&=\sqrt{\frac{(BC+AE)^{2}}{B^2D^2}-\frac{A}{B}}\\
&>\sqrt{\left(\frac{1}{2}+\frac{P^{e}_{0}}{2P^{e}_{1}\|\mathbf{g}\|^{2}}\right)^{2}-
\frac{P^{e}_{0}}{P^{e}_{1}\|\mathbf{g}\|^{2}}}\\
&=\left|\frac{1}{2}-\frac{P^{e}_{0}}{2P^{e}_{1}\|\mathbf{g}\|^{2}}\right|
\end{aligned}~.
\vspace*{-0.05in}
\end{equation}

Then, we can get:
\begin{equation}
\begin{aligned}
\rho^{*}_{+}&>\left(\frac{1}{2}+\frac{P^{e}_{0}}{2P^{e}_{1}\|\mathbf{g}\|^{2}}\right)+
\left|\frac{1}{2}-\frac{P^{e}_{0}}{2P^{e}_{1}\|\mathbf{g}\|^{2}}\right|\\
&=\left\{
\begin{array}{l}
1,\scriptsize{~~~~~~~~~~~}\frac{P^{e}_{0}}{2P^{e}_{1}\|\mathbf{g}\|^{2}}\leq\frac{1}{2}\\
\frac{P^{e}_{0}}{P^{e}_{1}\|\mathbf{g}\|^{2}},~\frac{P^{e}_{0}}{2P^{e}_{1}
\|\mathbf{g}\|^{2}}>\frac{1}{2}
\end{array}
\right.
\end{aligned}.
\vspace*{-0.05in}
\end{equation}

Note that $\rho^{*}_{+}\leq 1$, and hence
$\frac{P^{e}_{0}}{2P^{e}_{1}\|\mathbf{g}\|^{2}}$
should not be less than or equal to $\frac{1}{2}$.
However, when $\frac{P^{e}_{0}}{2P^{e}_{1}\|\mathbf{g}\|^{2}}>\frac{1}{2}$,
$\rho^{*}_{+}>\frac{P^{e}_{0}}{2P^{e}_{1}\|\mathbf{g}\|^{2}}$ is equivalent to $\rho^{*}_{+}>1$.
As a result, $\rho^{*}_{+}\notin (0,\rho_{max}]$ where
$\rho_{max}=\min\left(1,\sqrt{\frac{P^{e}_{0}}{P^{e}_{1}\|\mathbf{g}\|^2}}\right)$,
i.e., $\rho^{*}_{+}$ is not acceptable.

As for $\rho^{*}_{-}=\frac{(BC+AE)-\sqrt{\Delta}}{BD}$,
since $ABD^2>0$, $BC+AE>\sqrt{\Delta}$ holds, thus justifying $\rho^{*}_{-}>0$.
Then, we prove $\rho^{*}_{-}<\rho_{max}$ as follows.
First, we define a function $g(C)=(BC+AE)-\sqrt{\Delta}$.
Since the derivative of $g(C)$ to $C$,
$g'(C)=\frac{B\sqrt{\Delta}-B(BC+AE)}{\sqrt{\Delta}}<0$,
$g(C)$ is a monotonically decreasing function of variable $C$.
Let $C'=P^{e}_{T_{1}}|\chi|^{2}$,
then $C=\sigma^{2}_{n}+P^{e}_{1}|\chi|^{2}>C'$,
thus leading to $g(C)<g(C')$.
Similarly to the derivations of Eqs.~(13)--(15), we get:
\begin{equation}
\begin{aligned}
\rho^{*}_{-}=\frac{g(C)}{BD}
<\frac{g(C')}{BD}=\left\{
\begin{array}{l}
1,\scriptsize{~~~~~~~~~~~}\frac{P^{e}_{0}}{2P^{e}_{1}\|\mathbf{g}\|^{2}}\geq\frac{1}{2}\\
\frac{P^{e}_{0}}{P^{e}_{1}\|\mathbf{g}\|^{2}},~\frac{P^{e}_{0}}{2P^{e}_{1}
\|\mathbf{g}\|^{2}}<\frac{1}{2}
\end{array}.
\right.
\end{aligned}
%\vspace*{-0.05in}
\end{equation}
Eq.~(16) is equivalent to $\rho^{*}_{-}<\rho^{2}_{max}=\min\left(1,\frac{P^{e}_{0}}{P^{e}_{1}\|\mathbf{g}\|^{2}}\right)$.
Since $0<\rho^{2}_{max}\leq 1$, $\rho^{2}_{max}\leq \rho_{max}$ holds, thus proving $\rho^{*}_{-}<\rho_{max}$.

Finally, we prove that $\rho^{*}_{-}$ could achieve the maximum $\varphi_{0}$.
Since it can be proved that $h(\rho)=\frac{BD}{2}\rho^{2}-(BC+AE)\rho+\frac{AD}{2}$ is
a monotonically decreasing function of variable $\rho\in(0,\rho_{max}]$,
when $0<\rho<\rho^{*}_{-}$, we get $h(\rho)>h(\rho^{*})=0$.
Similarly, when $\rho^{*}_{-}<\rho<\rho_{max}$,
$h(\rho)<0$ can be derived.
Thus, $\rho^{*}_{-}$ corresponds to the maximum $\varphi_{0}$.
The optimal steering factor is calculated as $\rho^{*}=\rho^{*}_{-}$.

\section{Generalization of DIS}

\subsection{Generalized Number of Interferences}

So far, we have assumed that the MBS sends a single data stream to MUE,
i.e., only one interference is imposed on the PUE.
When multiple desired signals are sent from a MBS,
the proposed DIS can be extended as follows.
Since picocells are deployed within the coverage of a macrocell,
interferences from the other MBSs are negligible.
For clarity of presentation,
Fig.~3 shows a two-interference situation as an example,
where $\mathbf{i}_{1}$ and $\mathbf{i}_{2}$ are the interferences.
Only one desired signal is considered.
\begin{figure}[ht]
%\vspace*{-0.1in}
\graphicspath{/fig}
\centering
\includegraphics[width=0.27\textwidth]{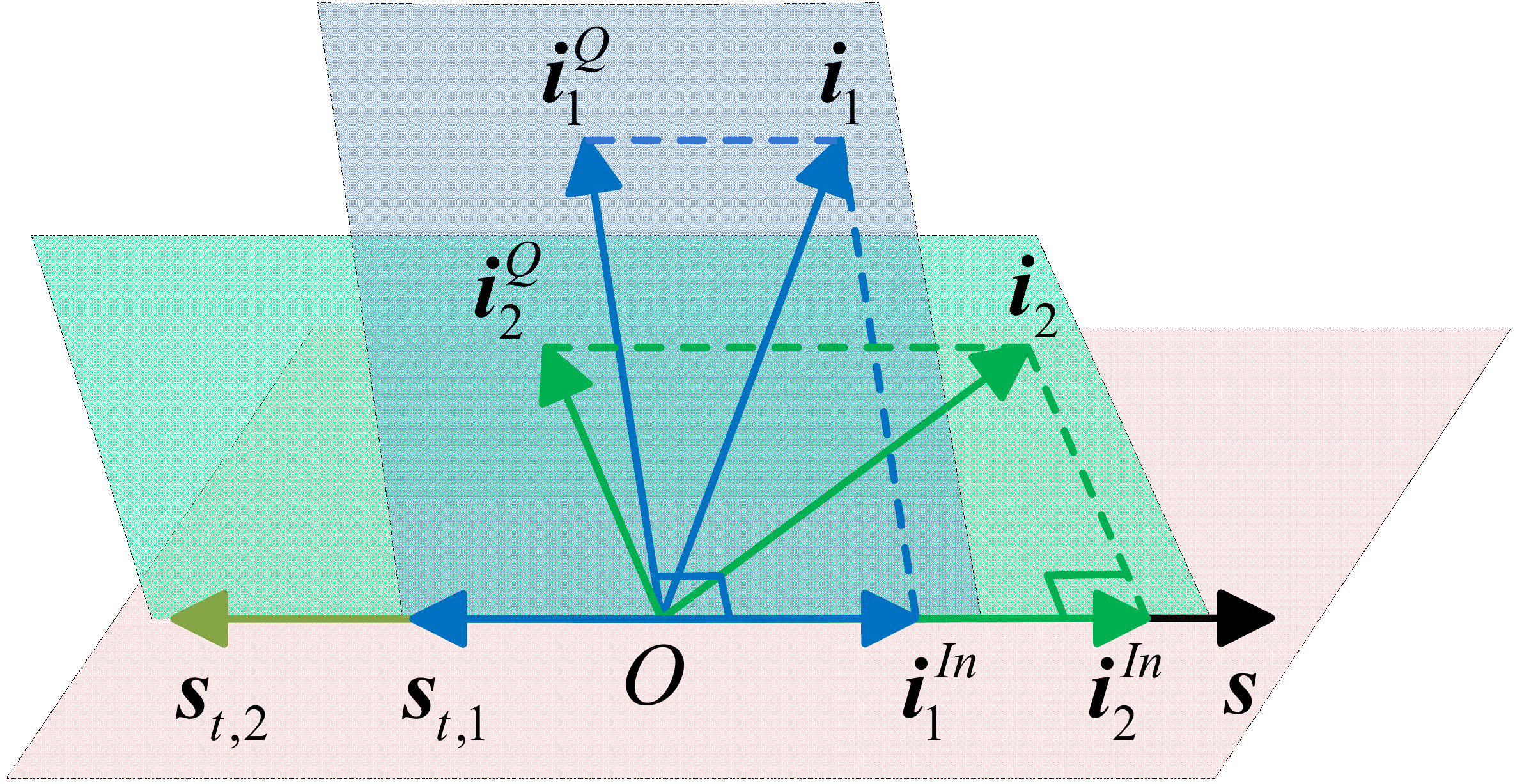}
\vspace{-3pt}
\caption{Generalization of the number of interferences.}
\label{fig: figure three}
\vspace*{-0.1in}
\end{figure}

As shown in this figure, each interference can be decomposed into an in-phase component
and a quadrature component w.r.t. the desired signal.
Then, DIS can be applied to each interference separately,
following the processing as described in Section III.

For $N>1$ interferences, the received signal at the victim PUE with DIS can be expressed as:
\begin{equation}
\vspace*{-0.05in}
\begin{aligned}
\mathbf{r}_{0}&=\sqrt{P^{e}_{0}-\sum^{N}_{n=1}P^{e}_{DIS,n}}\mathbf{H}_{0}\mathbf{p}_{0}x_{0}
+\sum^{N}_{n=1}\sqrt{P^{e}_{1,n}}\mathbf{H}_{10}\mathbf{p}_{1,n}x_{1,n}\\
&+\sum^{N}_{n=1}\sqrt{P^{e}_{DIS,n}}\mathbf{H}_{0}\mathbf{p}_{DIS,n}x_{1,n}+\mathbf{n}_{0}
\end{aligned},
\end{equation}
where $\mathbf{x}_{1}=[x_{1,1},\cdots,x_{1,n},\cdots,x_{1,N}]$ is the transmit data vector of MBS.
The transmission of $x_{1,n}$ causes the interference term
$\sqrt{P^{e}_{1,n}}\mathbf{H}_{10}\mathbf{p}_{1,n}x_{1,n}$ to the PUE,
in which $P^{e}_{1,n}$ is the transmit power for $x_{1,n}$, i.e., $P_{1,n}$,
incorporated with the path loss $10^{-0.1L_{10}}$.
$\mathbf{p}_{1,n}$ denotes the precoder for $x_{1,n}$.
$\sum^{N}_{n=1}P^{e}_{1,n}=P^{e}_{1}$ holds.
PBS generates a duplicate of this interference with the power overhead $P_{DIS,n}$
where $P^{e}_{DIS,n}=P_{DIS,n}10^{-0.1L_{0}}$
and the precoder $\mathbf{p}_{DIS,n}$ so as to adjust the interference
to an appropriate direction at the victim PUE.

Similarly to the derivation of Eq.~(5),
we can obtain the DIS design for the $n$-th interfering component as:
\begin{equation}
%\vspace*{-0.1in}
\left\{
\begin{array}{l}
\mathbf{p}_{DIS,n}=-\mathbf{H}^{-1}_{0}\mathbf{P}\mathbf{H}_{10}\mathbf{p}_{1,n}/
\|\mathbf{H}^{-1}_{0}\mathbf{P}\mathbf{H}_{10}\mathbf{p}_{1,n}\|\\
P^{e}_{DIS,n}=\rho^{2}_{n}P^{e}_{1,n}\|\mathbf{H}^{-1}_{0}\mathbf{P}\mathbf{H}_{10}\mathbf{p}_{1,n}\|^2
\end{array}
\right.,
%\vspace*{-0.05in}
\end{equation}
where $\mathbf{P}$ represents the projection matrix
depending only on the spatial feature of the desired transmission,
with which we can calculate the in-phase component of the interference
caused by the transmission of $x_{1,n}$ w.r.t. the intended signal.
$\rho_{n}$ is the steering factor for the steering signal carrying $x_{1,n}$.

One should note that when there are multiple interferences, it is difficult to
determine the optimal steering factors for all the interfering components.
However, we can allocate a power budget $P_{0,n}$,
satisfying $\sum^{N}_{n=1}P_{0,n}<P_{0}$, to each interference, and
then by applying DIS to each disturbance under its power budget constraint,
a vector of $n$ sub-optimal steering factors is achieved.
$P_{0,n}$ can be assigned with the same value or based on the strength of interferences.
The achievable SE of the PUE can then be calculated as:
\begin{equation}
c^{DIS}_{0}=\log_{2}\left\{
1+\frac{\left(P^{e}_{0}-\sum^{N}_{n=1}P^{e}_{DIS,n}\right)[\lambda^{(1)}_{0}]^{2}}{\sigma^{2}_{n}
+\sum^{N}_{n=1}I_{R,n}} \right\}
%\vspace*{-0.05in}
\end{equation}
where $I_{R,n}=P^{e}_{1}\|\mathbf{f}^{H}_{0}(1-\rho_{n})\mathbf{P}\mathbf{H}_{10}
\mathbf{p}_{1,n}\|^{2}$ ($n=1,\cdots,N$)
is the strength of the $n$-th residual interference to the intended transmission.
$\mathbf{f}_{0}=\mathbf{u}^{(1)}_{0}$ is the receive filter for data $x_{0}$.

To further elaborate on the extension of the proposed scheme,
we provide below an algorithm for $N=2$,
with which the optimal $\rho_{n}$ can be determined,
maximizing the system SE.
For simplicity, we use the function $f(\rho_{1},\rho_{2})$
to denote $c^{DIS}_{0}$ under $N=2$ in the following description.
The above algorithm can be extended to the case of $N>2$.
Due to space limitation, we do not elaborate on this any further in this paper.
\begin{algorithm}[!htb]%\small
\caption{\small{}}
\begin{algorithmic}[1]
%\REQUIRE~~\\
%\ENSURE~~\\
\STATE Take the derivative of $f(\rho_{1},\rho_{2})$ to $\rho_{1}$ and $\rho_{2}$ respectively,
to obtain $f'_{\rho_{n}}(\rho_{1},\rho_{2})=\frac{\partial f(\rho_{1},\rho_{2})}{\partial\rho_{n}}$ where $n=1,2$.
\STATE Compute the stationary point $(\tilde{\rho}_{1},\tilde{\rho}_{2})$ of $f(\rho_{1},\rho_{2})$ by solving the equation $f'_{\rho_{n}}(\rho_{1},\rho_{2})=0$. We define set $\Phi$ consisting of $(\tilde{\rho}_{1},\tilde{\rho}_{2})$.
\STATE Calculate the second-order derivative of $f(\rho_{1},\rho_{2})$
at the stationary point $(\tilde{\rho}_{1},\tilde{\rho}_{2})$, i.e.,
$f''_{\rho_{1},\rho_{2}}(\tilde{\rho}_{1},\tilde{\rho}_{2})=
\frac{\partial f(\rho_{1},\rho_{2})}{\partial\rho_{1} \partial\rho_{2}}\left|_{\rho_{1}=\tilde{\rho}_{1},\rho_{2}
=\tilde{\rho}_{2}}\right.$. For clarity of exposition, we define $\mathcal{A}=f''_{\rho_{1},\rho_{2}}(\tilde{\rho}_{1},\tilde{\rho}_{2})$.
Similarly, we define variables $\mathcal{B}=f''_{\rho_{1},\rho_{1}}(\tilde{\rho}_{1},\tilde{\rho}_{2})$ and
$\mathcal{C}=f''_{\rho_{2},\rho_{2}}(\tilde{\rho}_{1},\tilde{\rho}_{2})$.
\STATE Check whether the stationary point is an extreme point or not.
If $\mathcal{A}^{2}-\mathcal{BC}<0$ and $\mathcal{B}<0$, $(\tilde{\rho}_{1},\tilde{\rho}_{2})$ is an extreme point; otherwise not.
We can obtain the set of extreme points and the value of $f(\rho_{1},\rho_{2})$ at each extreme point correspondingly.
We define the extreme value set as $\Omega$.
\STATE Since both $\rho_{1}$ and $\rho_{2}$ range from $0$ to $1$,
i.e., $\rho_{n}\in(0,1]$,
the maximum value of $f(\rho_{1},\rho_{2})$ may exist at the boundary points.
We define set $\mathcal{F}=f(\rho_{1},\rho_{2})\left|_{\rho_{n}\in\{0,1\}, n=1,2}\right.$.
\STATE Determine the optimal $(\rho^*_{1},\rho^*_{2})$ outputting the maximum value of
$f(\rho_{1},\rho_{2})\left|_{\rho_{1}=\rho^*_{1},\rho_{2}=\rho^*_{2}}\right.$
by searching all the elements in sets $\Omega$ and $\mathcal{F}$.
\end{algorithmic}
\end{algorithm}
%\vspace*{-0.1in}

\subsection{Generalized Number of Desired Data Streams}

We now generalize the number of desired signals, denoted by $M$, sent from PBS to its PUE.
For clarity of exposition, we take $M=2$
and the number of interferences $N=1$ as an example as shown in Fig.~4.
However, this discussion can be readily extended to more general parameter settings.
%The spatial features of multiple desired signals and interferences observed by the PUE
%are illustrated by Fig.~4.
As can be seen from the figure, the interference $\mathbf{i}$ forms
a plane with each of the desired signals $\mathbf{s}_{m}$ where $m=1,2$.
The projection of $\mathbf{i}$ onto $\mathbf{s}_{m}$ is denoted by $\mathbf{i}^{In}_{m}$
(in-phase component), whereas the quadrature component is $\mathbf{i}^{Q}_{m}$.
By applying DIS to each $(\mathbf{i},\mathbf{s}_{m})$ pair,
a set of steering signals can be determined.
\begin{figure}[!htb]
\vspace*{-0.1in}
\graphicspath{/fig}
\centering
\includegraphics[width=0.27\textwidth]{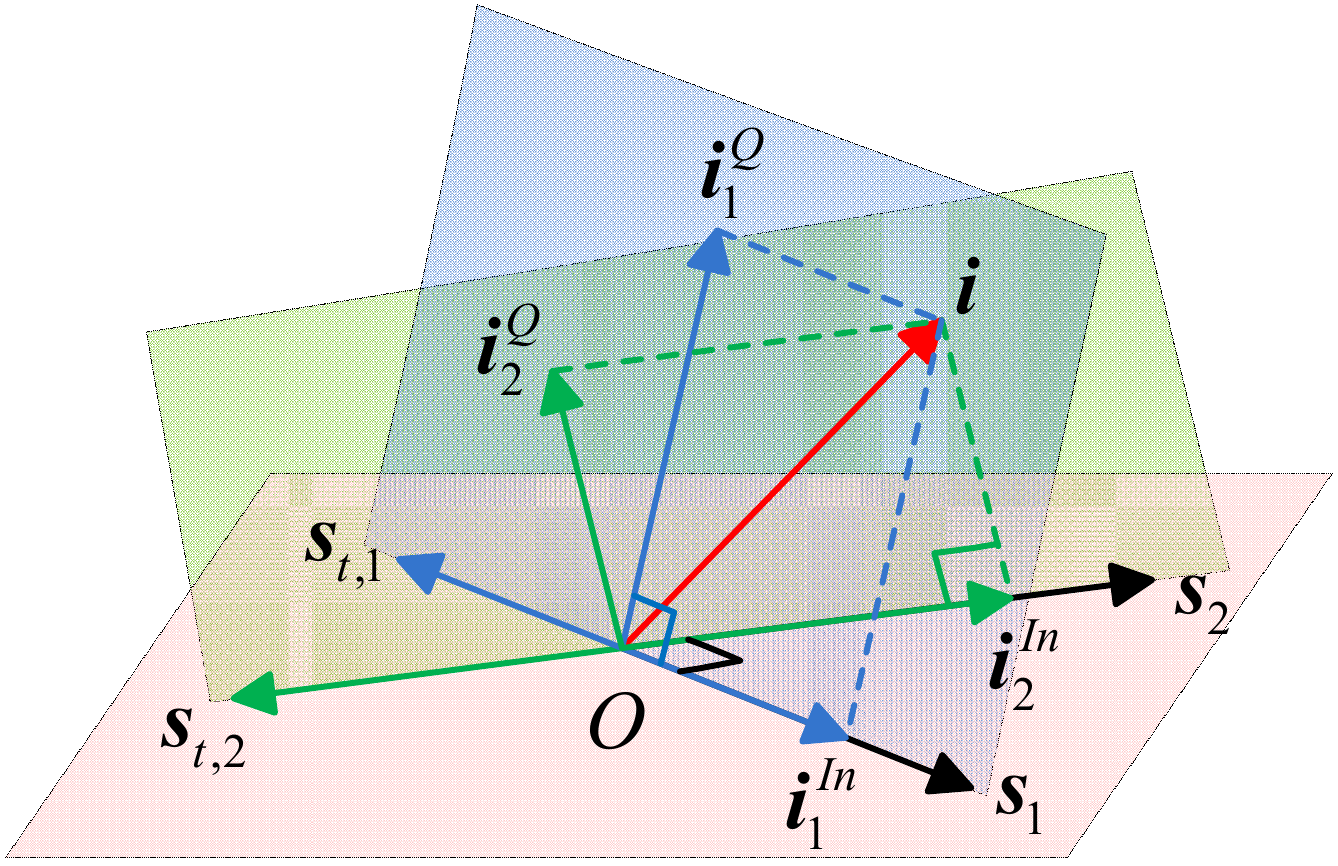}
\vspace{-3pt}
\caption{Generalization of the number of desired signals.}
\label{fig: figure four}
\vspace*{-0.05in}
\end{figure}

Since multiple data streams are sent from PBS to PUE via mutually orthogonal eigenmodes/subchannels,
and the steering signal is opposite to the spatial feature of
the desired transmission it intends to protect,
an arbitrary steering signal, say $\mathbf{s}_{t,m}$,
is orthogonal to any of the other desired signals $\mathbf{s}_{m'}$ where $m'\neq m$.
Hence, no additional interference will be created by the steering signal $\mathbf{s}_{t,m}$.

Based on the above discussion,
the number of steering signals is equal to that of the desired data streams $M$.
Thus, the mixed signal received at PUE can be expressed as:
\vspace*{-0.1in}
\begin{equation}
\begin{aligned}
\mathbf{r}_{0}&=\sum^{M}_{m=1}\sqrt{P^{e}_{0,m}-P^{e}_{DIS,m}}\mathbf{H}_{0}\mathbf{p}_{0,m}x_{0,m}
+\sqrt{P^{e}_{1}}\mathbf{H}_{10}\mathbf{p}_{1}x_{1}\\
&+\sum^{M}_{m=1}\sqrt{P^{e}_{DIS,m}}\mathbf{H}_{0}\mathbf{p}_{DIS,m}x_{1}+\mathbf{n}_{0}
\end{aligned}
%\vspace*{-0.1in}
\end{equation}
where $\mathbf{x}_{0}=[x_{0,1},\cdots,x_{0,m},\cdots,x_{0,M}]$ is the transmit data vector of PBS.
$P^{e}_{0,m}$ denotes the transmit power budget for $x_{0,m}$, i.e., $P_{0,m}$,
incorporated with the path loss $10^{-0.1L_{0}}$.
$P_{DIS,m}$ is the power cost for steering the interference
away from the $m$-th desired signal, and $P^{e}_{DIS,m}=P_{DIS,m}10^{0.1L_{0}}$.
$\mathbf{p}_{0,m}$ and $\mathbf{p}_{DIS,m}$ represent the precoders for
$x_{0,m}$ and its steering signal, respectively.
Similarly to the derivation of Eqs.~(5) and (18), the
DIS solution for multi-desired-signal situation can be readily obtained.
In such a case, the achievable SE of PUE can be expressed as:
\begin{equation}
c^{DIS}_{0}=\sum^{M}_{m=1}\log_{2}\left\{
1+\frac{\left(P^{e}_{0,m}-P^{e}_{DIS,m}\right)[\lambda^{(m)}_{0}]^{2}}{\sigma^{2}_{n}+I_{R,m}}
\right\}.
\vspace*{-0.05in}
\end{equation}
\\
$\lambda^{(m)}_{0}$ denotes the amplitude gain of the $m$-th desired transmission from PBS to PUE.
The residual interference $I_{R,m}=P^{e}_{1}\|\mathbf{f}^{H}_{0,m}(1-\rho_{m})\mathbf{P}_{m}\mathbf{H}_{10}
\mathbf{p}_{1}\|^{2}$
where $\mathbf{f}_{0,m}=\mathbf{u}^{(m)}_{0}$ is the receive filter for data $x_{0,m}$.
The projection matrix
$\mathbf{P}_{m}=\mathbf{d}_{\mathbf{s},m}(\mathbf{d}^{T}_{\mathbf{s},m}\mathbf{d}_{\mathbf{s},m})^{-1}
\mathbf{d}^{T}_{\mathbf{s},m}$
where $\mathbf{d}_{\mathbf{s},m}=\frac{\mathbf{H}_{0}\mathbf{p}_{0,m}}{\|\mathbf{H}_{0}\mathbf{p}_{0,m}\|}$.

It should be noted that the optimal $P^{e}_{DIS,m}$, or equivalently $\rho_{m}$ is dependent on $P^{e}_{0,m}$
where $\sum^{M}_{m=1}P^{e}_{0,m}=P^{e}_{0}$ holds.
Thus, different power allocations will yield different DIS solutions.
For example, $P_{0}$ can be equally allocated to the $M$ intended data transmissions,
or in terms of the quality of subchannels and/or
the strength of interference imposed on each desired signal.
Then, suboptimal performance w.r.t. the transmission from PBS to PUE is achieved.
How to jointly determine the optimal $P_{0,m}$ and $P_{DIS,m}$ is our future work.
%For conciseness, we do not elaborate here.
%For simplicity, in our evaluation (see in Section V)
%we assume $P_{0}$ is equally allocated to $M$ intended data transmissions,
%yielding suboptimal results.

\subsection{Generalized Number of PBSs and PUEs}

We discuss the generalization of the number of PBSs deployed in the coverage of a macrocell
and the number of PUEs served by each PBS.
As mentioned before, PBSs are installed by the network operator.
Inter-picocell interference could therefore be effectively avoided by
the operator's planned deployment or resource allocation.
Even when inter-picocell interference exists,
our scheme can be directly applied by
treating the interfering PBS as the MBS in this paper,
and DIS is implemented at the PBS associated with the victim PUE.
It should be noted that the proposed DIS is applicable to
the scenario with asymmetric interferences.
Otherwise, concurrent data transmissions should be scheduled
or other schemes should be adopted to address the interference problem,
which is beyond the scope of this paper.
As for the multi-PUE case, each PUE can be assigned an exclusive channel so as to avoid
co-channel interference, which is consistent with various types of wireless communication systems, such as WLANs.
In summary, with an appropriate system design, the proposed DIS can be applied to the system with multiple PBSs and PUEs.

\section{Evaluation}
\setcounter{figure}{7}
\begin{figure*}
%\vspace*{-0.05in}
\centering
\graphicspath{/fig}
\begin{minipage}{0.66\columnwidth}
\centering
\subfigure[$N_{T_{0}}$ is variable.]{
%\label{fig:subfig:a} %% label for first subfigure
\includegraphics[width=1\columnwidth, height = 1.75in]{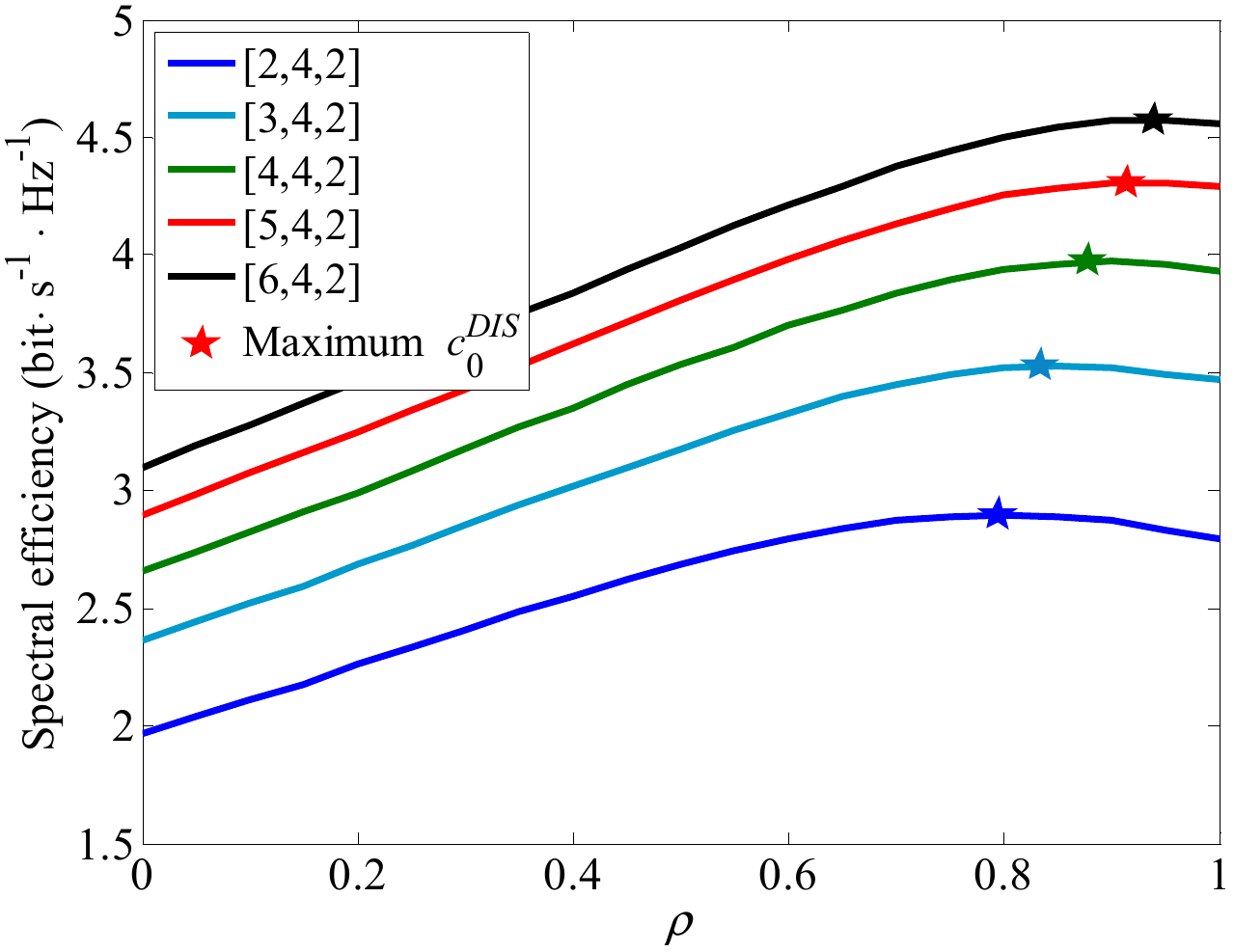}}
\vspace{-5pt}
\end{minipage}
\hfill
\begin{minipage}{0.66\columnwidth}
\centering
\subfigure[$N_{T_{1}}$ is variable.]{
%\label{fig:subfig:b} %% label for second subfigure
\includegraphics[width=1\columnwidth, height = 1.75in]{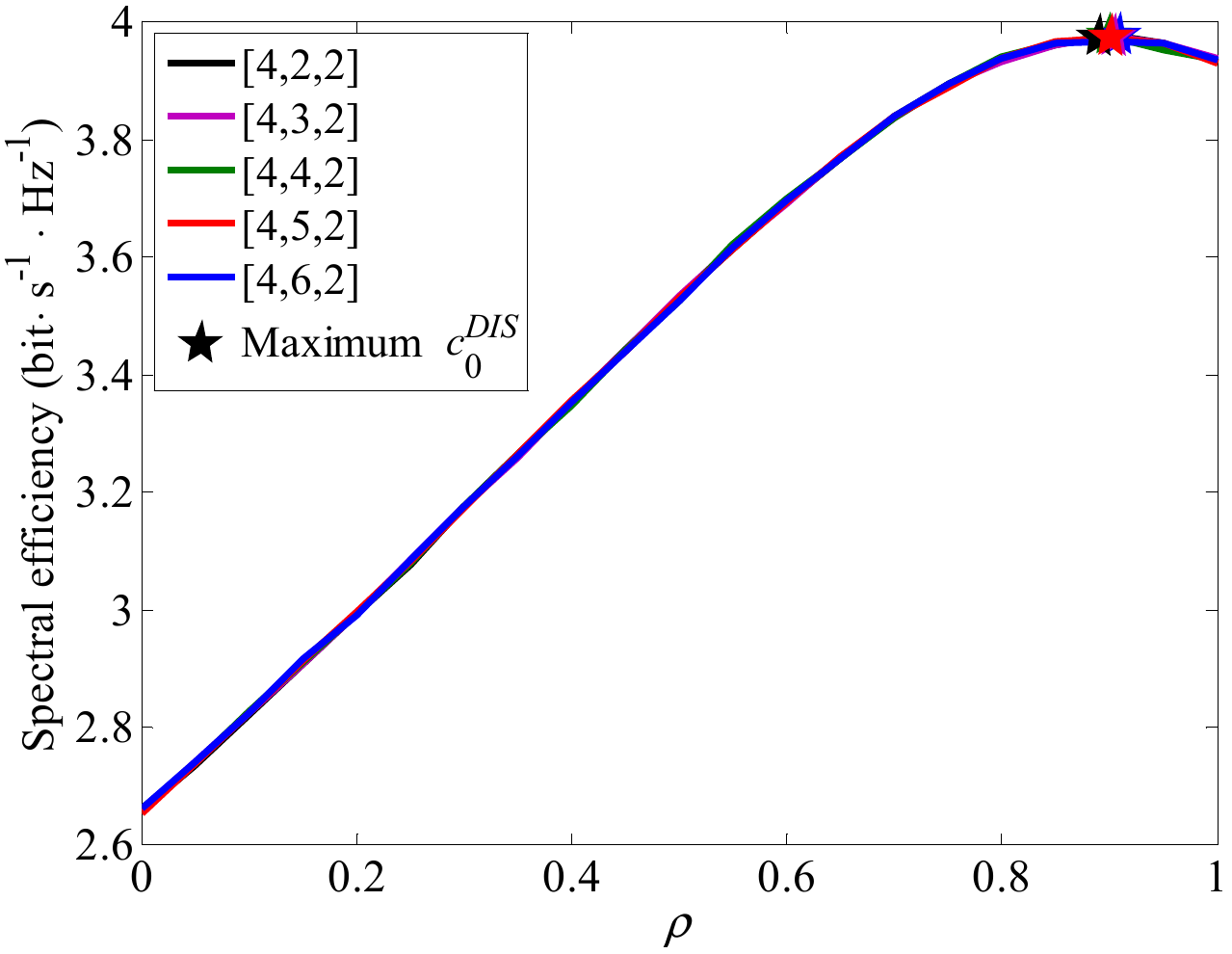}}
\vspace{-5pt}
%\label{fig:subfig:b} %% label for entire figure \end{figure}
\end{minipage}
\hfill
\begin{minipage}{0.66\columnwidth}
\centering
\subfigure[$N_{R_{0}}$ is variable.]{
%\label{fig:subfig:c} %% label for second subfigure
\includegraphics[width=1\columnwidth, height = 1.75in]{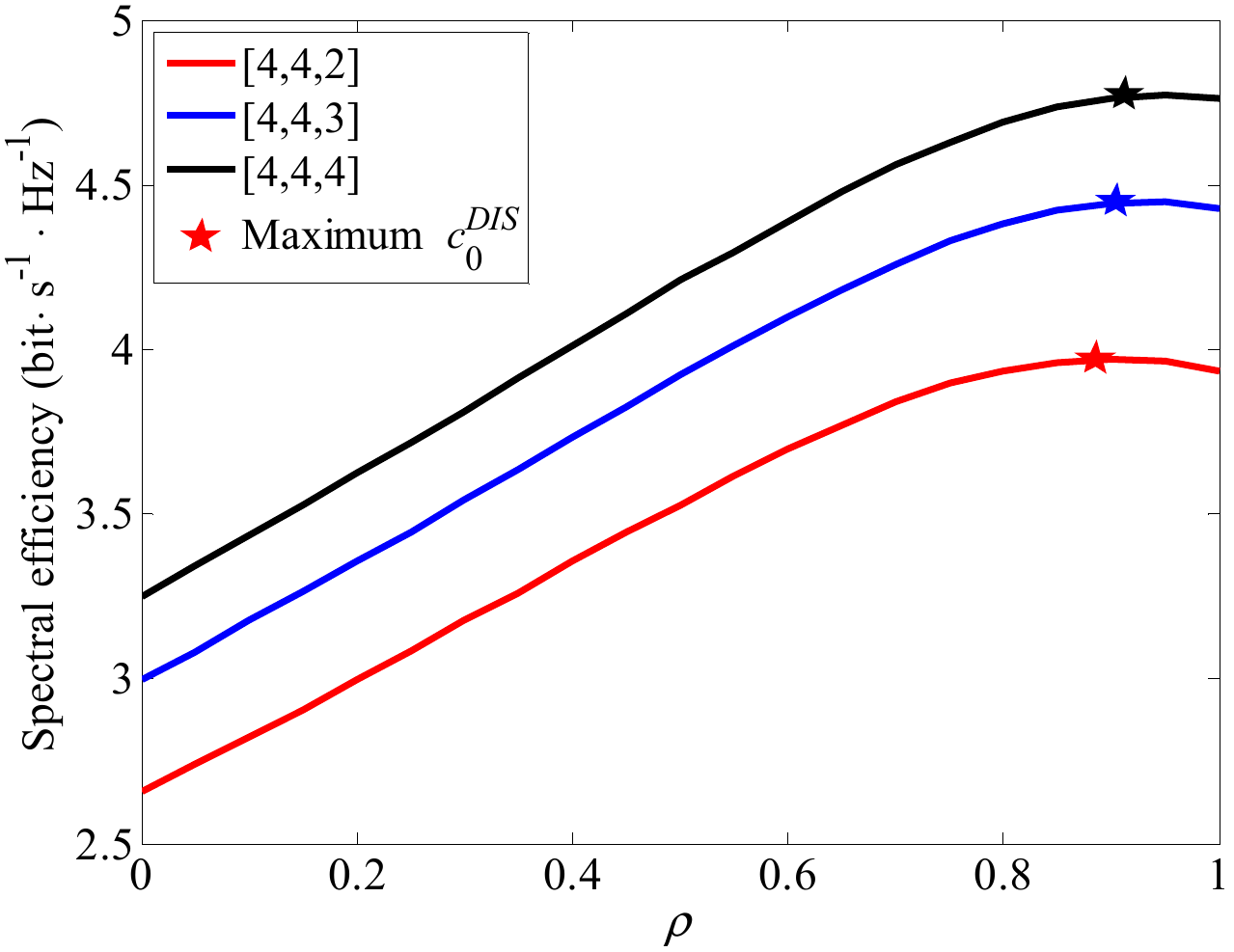}}
\vspace{-5pt}
\end{minipage}
\caption{SE of PUE vs.~$\rho$ under $\bar{\gamma}=5dB$, $\xi=1$, $M=N=1$, and different antenna settings.}
\vspace*{-0.2in}
\end{figure*}

We evaluate the performance of the proposed mechanism using MATLAB.
We set $d=300m$, $D=3000m$, $P_{0}=23dBm$ and $P_{1}=46dBm$ [14].
%Beamforming is adopted by the transmission in picocell, i.e., $M=1$.
The path loss is set to $L_{10}=128.1+37.6\log_{10}(\eta_{10}/10^{3})~dB$ and
$L_{0}=38+30\log_{10}(\eta_{0})~dB$
where $\eta_{0}\leq d$ and $\eta_{10}\leq D$.
Since $L_{0}$ and $L_{10}$ are dependent on the network topology,
$P^{e}_{0}$ ranges from $-89dBm$ to $23dBm$,
whereas $P^{e}_{1}$ varies between $-100dBm$ to $46dBm$.
For clarity of presentation, we adopt $\bar{\gamma}=10lg(\gamma)$
where $\gamma=P^{e}_{1}/\sigma^{2}_{n}$.
We also define $\xi=P^{e}_{0}/P^{e}_{1}$.
Then, based on the above parameter settings, $\xi\in[-135,123]~dB$.
Note, however, that we obtained this result for extreme boundary situations,
making its range too wide to be useful.
Without specifications,
the simulation is done under $N_{T_{0}}=N_{T_{1}}=N_{R_{0}}=2$ antenna configuration.
However, same conclusion can be drawn with various parameter settings.
In practice, a PBS should not be deployed close to MBS and
mobile users may select an access point based on
the strength of reference signals from multiple access points.
Considering this practice, we set $\xi\in [0.1,100]$ in our simulation.
There are $M$ desired signals and $N$ interferences.
In the following simulation, when the power overhead of an IM scheme exceeds $P_{0}$ at the victim Tx,
we simply switch to non-IM mode, i.e., matched filtering (MF) is employed by letting $\mathbf{f}_{0,m}=\mathbf{u}^{(m)}_{0}$,
while the interference remains unchanged.
\setcounter{figure}{4}
\begin{figure}[!htb]
\vspace{-0.1in}
\graphicspath{/fig}
\centering
\subfigure[Weak interference.]{
\includegraphics[width=1.67in,height=1.3in]{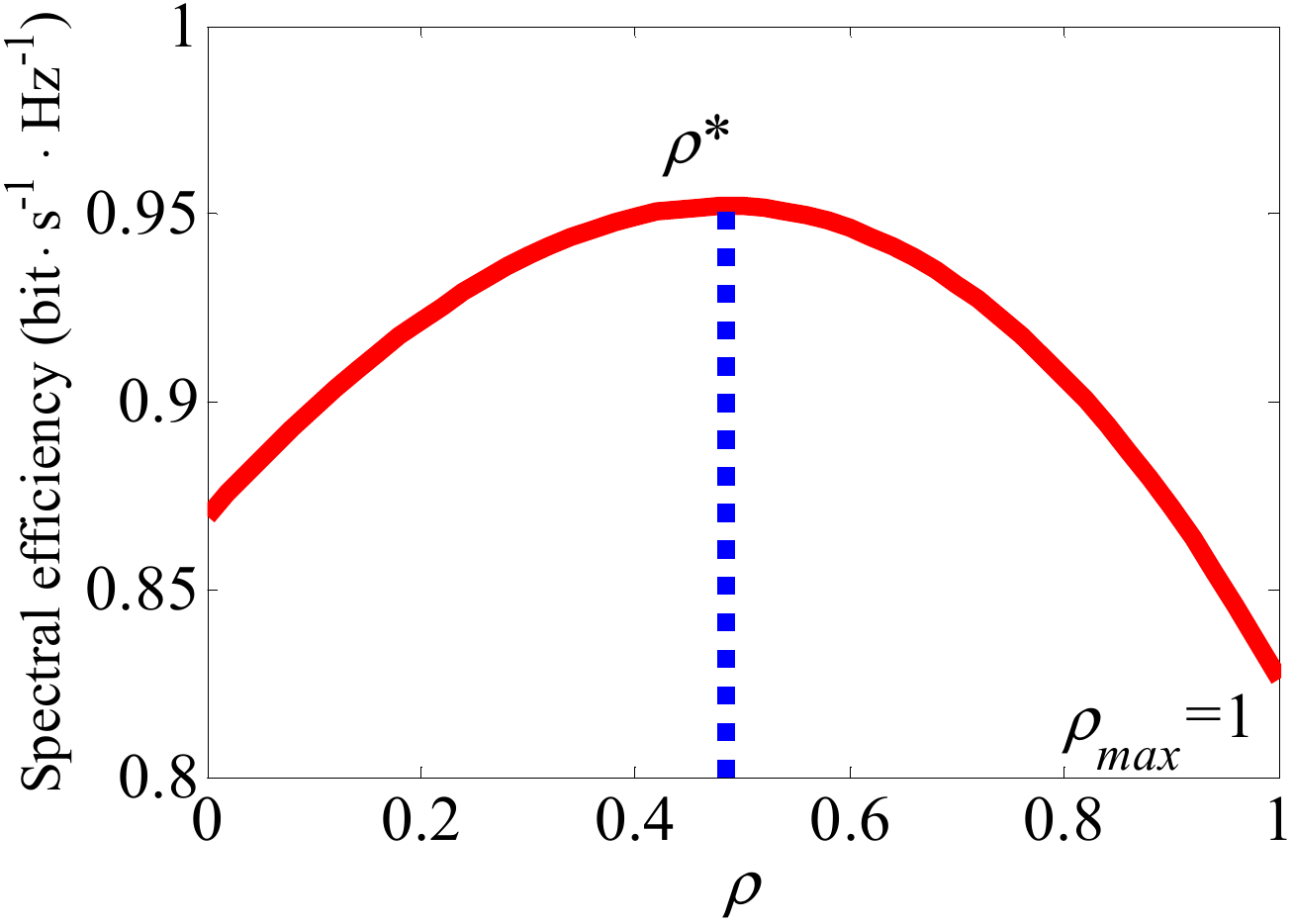}}
\subfigure[Strong interference.]{
\includegraphics[width=1.67in,height=1.28in]{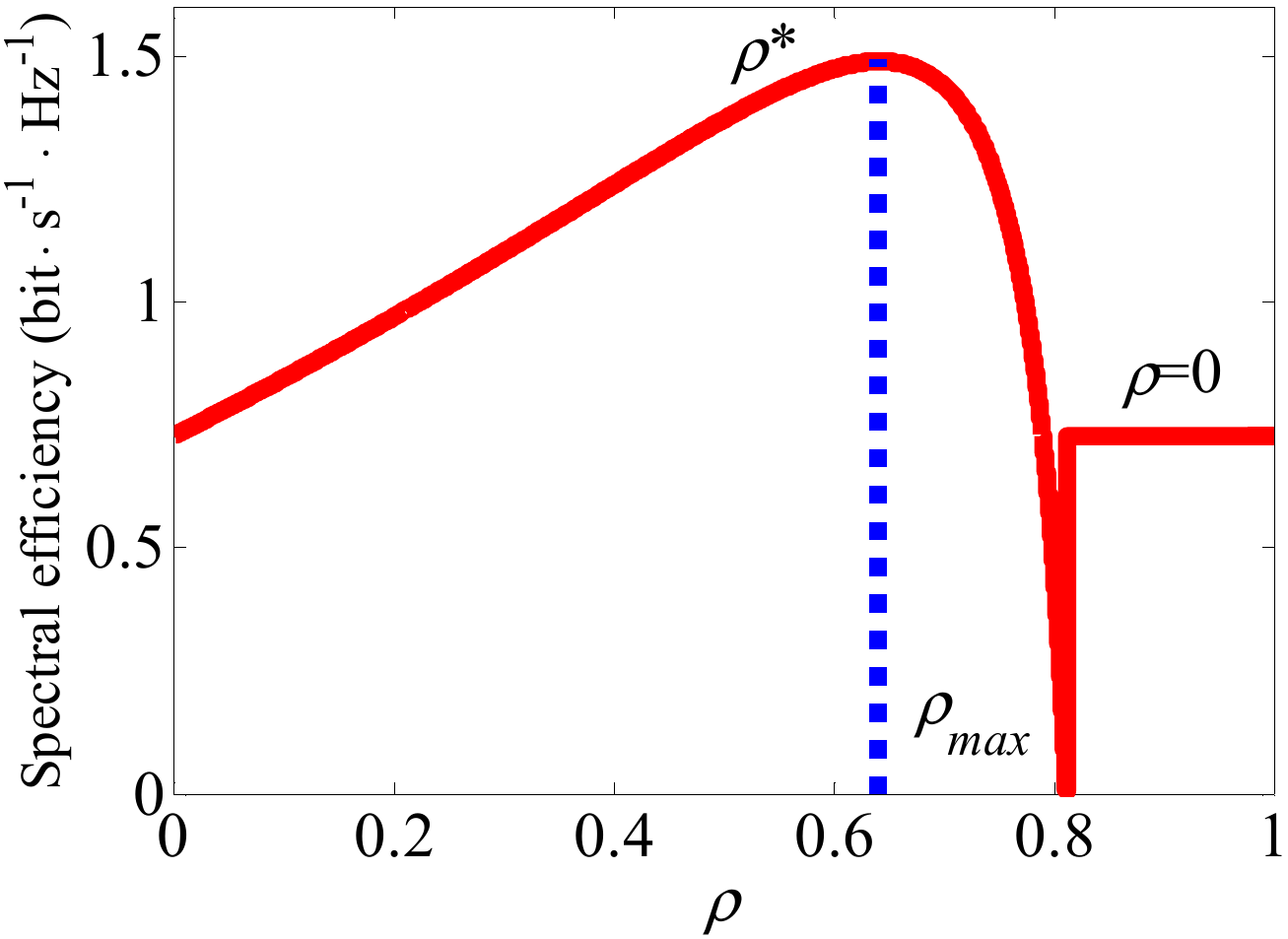}}
\vspace{-10pt}
\caption{SE of PUE vs. $\rho$ under $\bar{\gamma}=0dB$, $M=N=1$, and $\xi=1$.}
\label{fig: figure five} %% label for entire figure
\vspace*{-0.1in}
\end{figure}
%\begin{figure}[!htb]
%\graphicspath{/fig}
%\centering
%\subfigure[Weak interference.]{
%\includegraphics[width=0.4\textwidth,height=0.3\textwidth]{fig/figure5a.pdf}}
%\hspace{0.5in}
%\subfigure[Srong interference.]{
%\includegraphics[width=0.4\textwidth,height=0.3\textwidth]{fig/figure5b.pdf}}
%\caption{SE of PUE vs. $\rho$ under $N_{T_{0}}=N_{T_{1}}=N_{R_{0}}=2$,
%$\bar{\gamma}=0dB$, and $\xi=1$.}
%\label{fig: figure five} %% label for entire figure
%%\vspace*{-0.1in}
%\end{figure}

Fig.~5 shows two samples of the relationship between $\rho$ and PUE's achievable SE.
The interference shown in Fig.~5(a) is relatively weak,
and hence the transmit power of PBS is sufficient for OIS,
i.e., $\rho_{max}$ can be as large as $1$.
In Fig.~5(b), since the interference is strong, and hence,
when $\rho>\rho_{max}$ where $\rho_{max}<1$,
there won't be enough power for PBS to realize OIS.
In such a case, we simply switch off IS and adopt non-IM.
In both figures, the optimal $\rho$, denoted by $\rho^{*}$,
is computed as in Section III, which corresponds to the maximum SE.
We can conclude from Fig.~5 that in order to better utilize
the transmit power for both IS and data transmission,
it is necessary to intelligently determine the appropriate strength of the steering signal,
or equivalently, the direction into which the interference is steered.
\setcounter{figure}{5}
\begin{figure}[ht]
%\vspace*{-0.1in}
\graphicspath{/fig}
\centering
\includegraphics[width=0.33\textwidth]{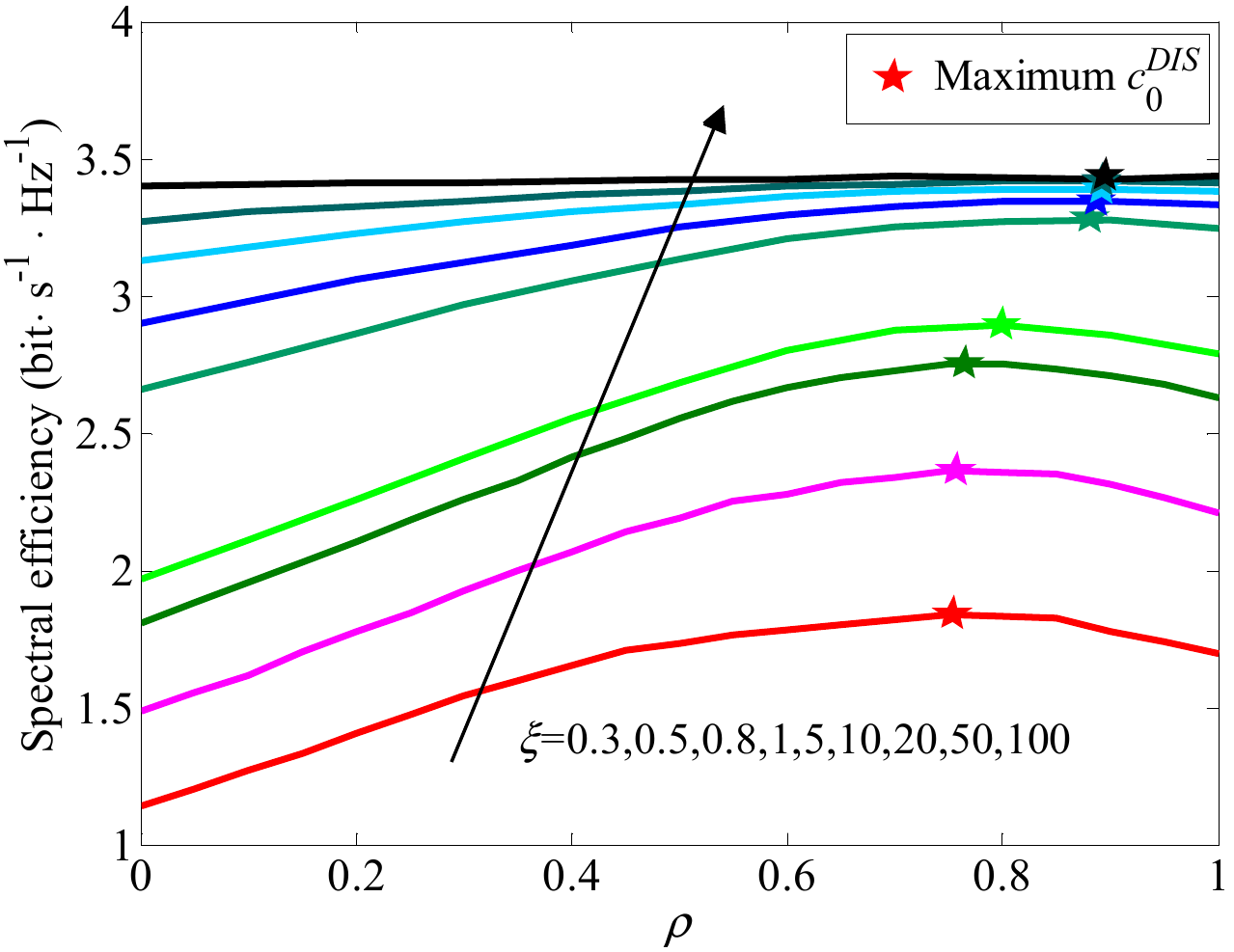}
\vspace{-5pt}
\caption{SE of PUE vs. $\rho$ under $\bar{\gamma}=5dB$, $M=N=1$, and different $\xi$.}
\label{fig: figure six}
\vspace*{-0.05in}
\end{figure}

Fig.~6 plots the PUE's average SE versus $\rho$ for different $\xi$.
The average $\rho^{*}$, marked by pentagram, which corresponds to the PUE's maximum SE,
grows as $\xi$ increases. When $\xi$ gets too high,
a large portion of interference is preferred to be mitigated.
As shown in the figure, given $\xi=100$, the average $\rho^{*}$ is approximately $0.9$.
In addition, since the strength of the desired signal relative to the interference
grows with an increase of $\xi$, the PUE's SE performance improves with $\rho$.

Fig.~7 shows the average $\rho^{*}$ versus $\bar{\gamma}$ under
%$N_{T_{0}}=N_{T_{1}}=N_{R_{0}}=2$,
$M=N=1$ and different $\xi$. As the figure shows, with fixed $\xi$,
the average $\rho^{*}$ grows with an increase of $\bar{\gamma}$.
This is because the interference gets stronger with increasing $\bar{\gamma}$,
and hence, to achieve the maximum SE, $\rho^{*}$ should increase,
i.e., more interference imposing onto the desired signal should be mitigated.
Given the same $\bar{\gamma}$,
the average $\rho^{*}$ grows with an increase of $\xi$,
which is consistent with Fig.~6.
\setcounter{figure}{6}
\begin{figure}[ht]
%\vspace*{-0.05in}
\graphicspath{/fig}
\centering
\includegraphics[width=0.33\textwidth]{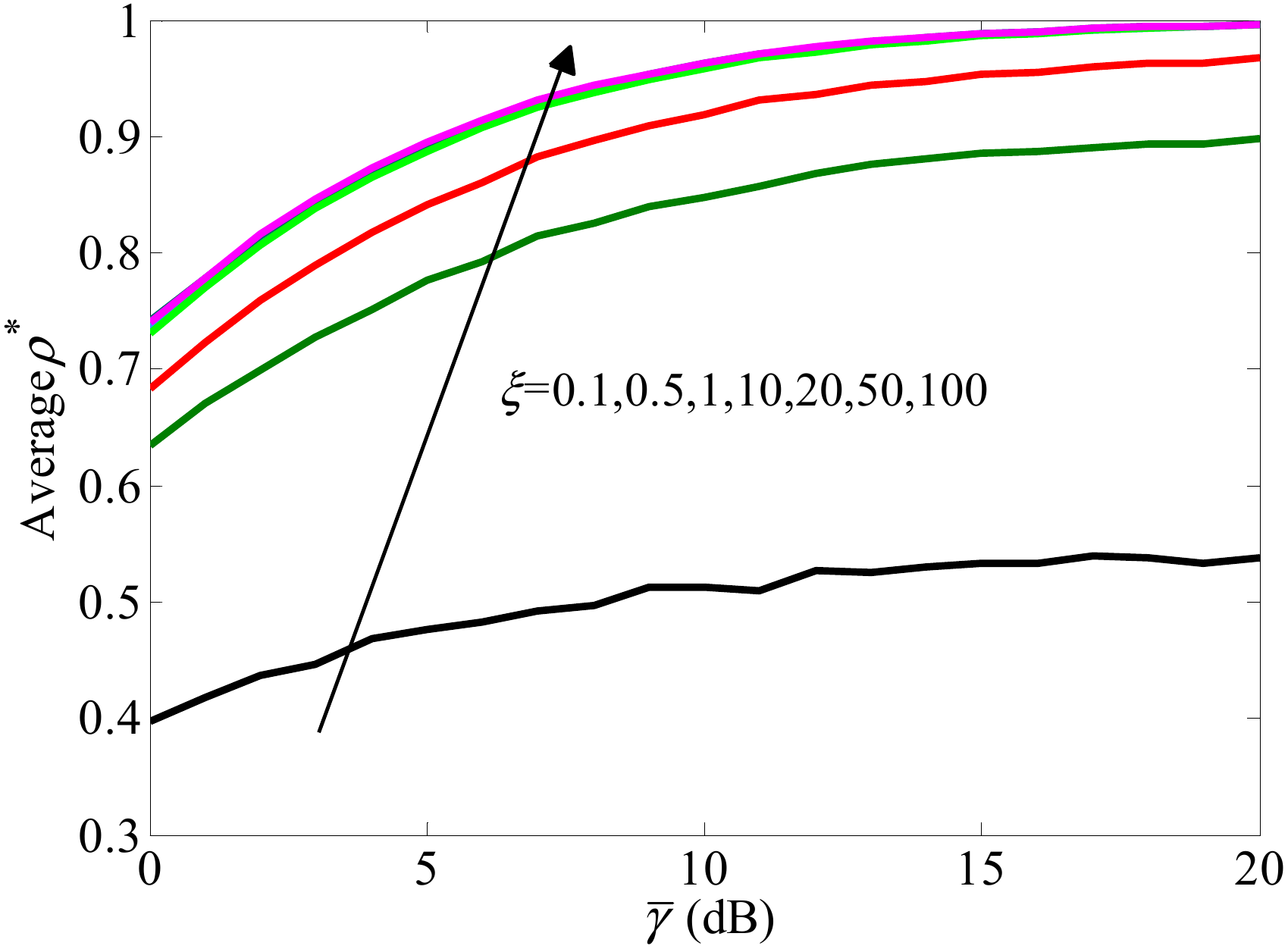}
\vspace{-5pt}
\caption{Average $\rho^{*}$ vs. $\bar{\gamma}$ under $M=N=1$, and different $\xi$.}
\label{fig: figure seven}
\vspace*{-0.05in}
\end{figure}
%\begin{figure}[!htb]
%\graphicspath{/fig}
%\centering
%\subfigure[$N_{T_{0}}$ is variable.]{
%\includegraphics[width=0.4\textwidth,height=0.3\textwidth]{fig/figure8b.pdf}}
%\hspace{0.5in}
%\subfigure[$N_{T_{1}}$ is variable.]{
%\includegraphics[width=0.4\textwidth,height=0.3\textwidth]{fig/figure8a.pdf}}
%\hspace{0.5in}
%\subfigure[$N_{R_{0}}$ is variable.]{
%\includegraphics[width=0.4\textwidth,height=0.3\textwidth]{fig/figure8c.pdf}}
%\caption{SE of PUE vs.~$\rho$ under $\bar{\gamma}=5dB$, $\xi=1$, and different antenna settings.}
%\label{fig: figure eight} %% label for entire figure
%%\vspace*{-0.1in}
%\end{figure}
\setcounter{figure}{8}

Fig.~8 plots PUE's average SE along with $\rho$ under different antenna settings.
We use a general form $[N_{T_{0}}~N_{T_{1}}~N_{R_{0}}]$ to express the antenna configuration.
In Fig.~8(a), $N_{T_{1}}$ and $N_{R_{0}}$ are fixed,
and $N_{T_{0}}$ varies from $2$ to $6$.
Since the transmit array gain of the desired signal grows with an increase of $N_{T_{0}}$,
meaning that the desired signal becomes stronger relative to the interference,
and more interference can be eliminated by DIS with the same power overhead as $N_{T_{0}}$ grows,
both the average $\rho^{*}$ and the achievable SE improve as $N_{T_{0}}$ increases.
In Fig.~8(b), $N_{T_{0}}$ and $N_{R_{0}}$ are
fixed while $N_{T_{1}}$ ranges from $2$ to $6$.
Although $N_{T_{1}}$ varies, MBS causes random interferences to PUE as
the PUE adopts $\mathbf{f}_{0}$ to decode $x_{0}$ regardless of the interference channel $\mathbf{H}_{10}$.
Hence, both $\rho^{*}$ and the PUE's average SE under different $N_{T_{1}}$ remain similar.
In Fig.~8(c), $N_{T_{0}}$ and $N_{T_{1}}$ are fixed
while $N_{R_{0}}$ ranges from $2$ to $4$.
With such antenna settings, since $N_{T_{0}}$ and $N_{T_{1}}$ are fixed,
the processing gain with the transmit antenna array doesn't change
for the desired signal or the interference.
However, as $N_{R_{0}}$ increases, the receive gain for the intended signal grows as
the filter vector $\mathbf{f}_{0}$, an $N_{R}\times 1$ vector designed to match $\mathbf{H}_{0}$.
As a result, the desired signal, relative to the interference,
after the receive filtering becomes stronger with an increase of $N_{R_{0}}$,
thus enhancing the PUE's SE.
\begin{figure}[ht]
%\vspace*{-0.05in}
\graphicspath{/fig}
\centering
\includegraphics[width=0.33\textwidth]{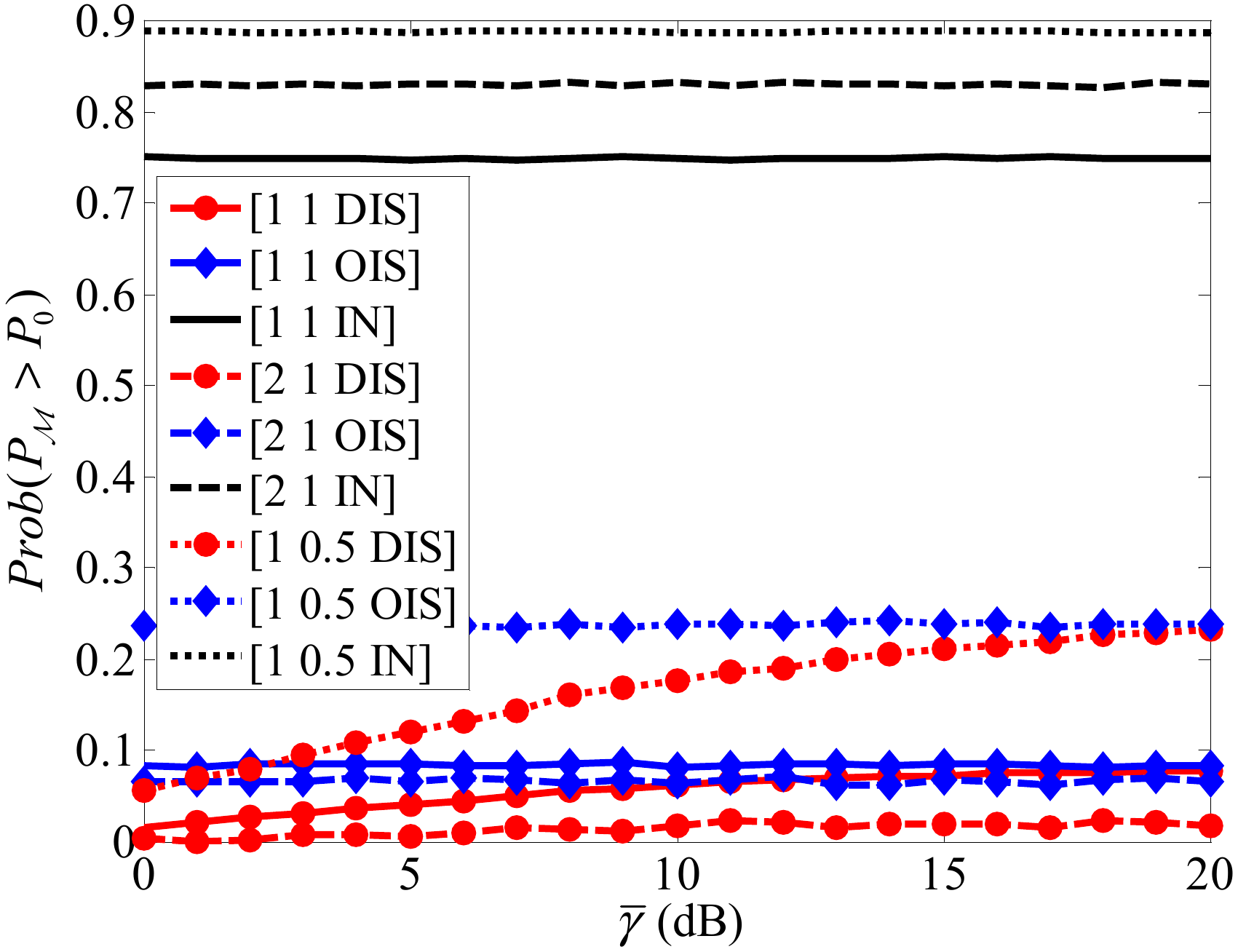}
\vspace{-5pt}
\caption{$Prob(P_{\mathcal{M}}>P_{0})$ vs. $\bar{\gamma}$ under $M=1$, and different $\xi$ and $N$.}
\label{fig: figure nine}
%\vspace*{-0.05in}
\end{figure}

Fig.~9 shows the probabilities that the power overhead of DIS,
OIS and IN are greater than $P_{0}$, i.e., $Prob(P_{\mathcal{M}}>P_{0})$
where $P_{\mathcal{M}}$ denotes the power cost at PBS with IM scheme $\mathcal{M}$.
We use a general form $[N~\xi~\mathcal{M}]$ to denote the parameter settings for different
mechanisms, where $N$ is the number of interferences.
$\xi=P^{e}_{0}/P^{e}_{1}$.
$P^{e}_{1}$ is the total power of the interferer, i.e., $P_{1}$, incorporated with path loss.
$Prob(P_{\mathcal{M}}>P_{0})$ is shown to increase as $\xi$ decreases
since a small $\xi$ results in a strong interference, incurring higher $P_{\mathcal{M}}$.
$Prob(P_{IN}>P_{0})$ is notably higher than
$Prob(P_{OIS}>P_{0})$ and $Prob(P_{DIS}>P_{0})$,
and $Prob(P_{DIS}>P_{0})$ is less than $Prob(P_{OIS}>P_{0})$.
This is because IN consumes more power than DIS and OIS.
For DIS, the power overhead increases as $\bar{\gamma}$ grows and
approaches OIS when $\bar{\gamma}$ becomes too large.
One may note in Fig.~9 that with fixed $\xi$ and $P^{e}_{1}$,
$Prob(P_{IN}>P_{0})$ with $N=2$ interferences is higher than
that with a single interference. However, as for DIS and OIS,
a larger $N$ produces lower $Prob(P_{\mathcal{M}}>P_{0})$.
This phenomenon can be explained by the results illustrated in Fig.~10.
\begin{figure*}[!htb]
%\vspace*{-0.1in}
\centering
\begin{minipage}{0.66\columnwidth}
\includegraphics[width=1\columnwidth, height = 1.75in]{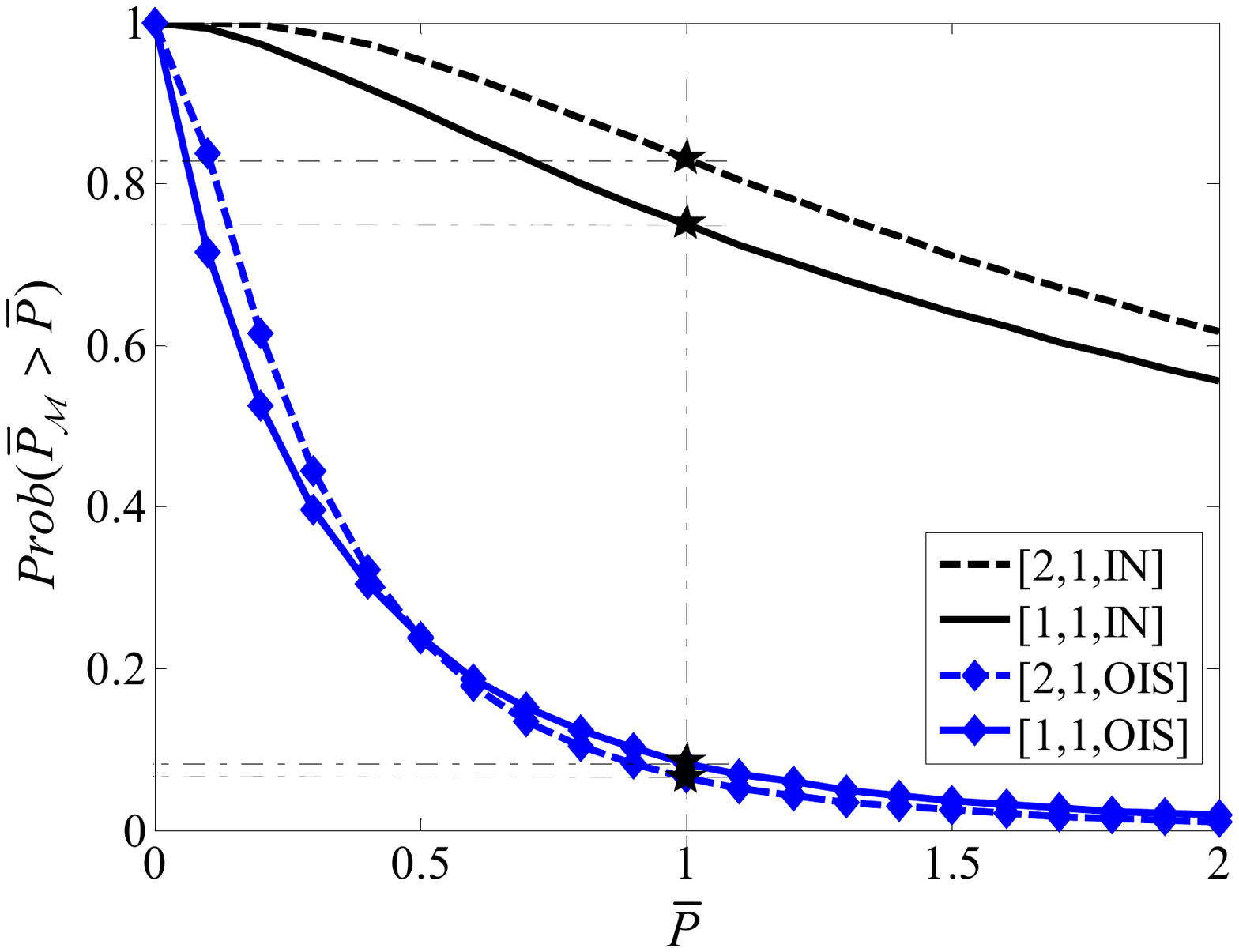}
\vspace{-16pt}
\caption{\fontsize{7.2pt}{\baselineskip}\selectfont{$Prob(P_{\mathcal{M}}>\bar{P})$ vs. $\bar{P}$ under $\xi=1$, $M=1$, and different $N$.}}
%\caption{$Prob(P_{\mathcal{M}}>\bar{P})$ vs. $\bar{P}$ under $N_{T_{0}}=N_{T_{1}}=N_{R_{0}}=2$, $\xi=1$, $M=1$, and different $N$.}
\label{fig: figure ten}
\end{minipage}
\hfill
\begin{minipage}{0.66\columnwidth}
\centering
\includegraphics[width=1\columnwidth, height = 1.75in]{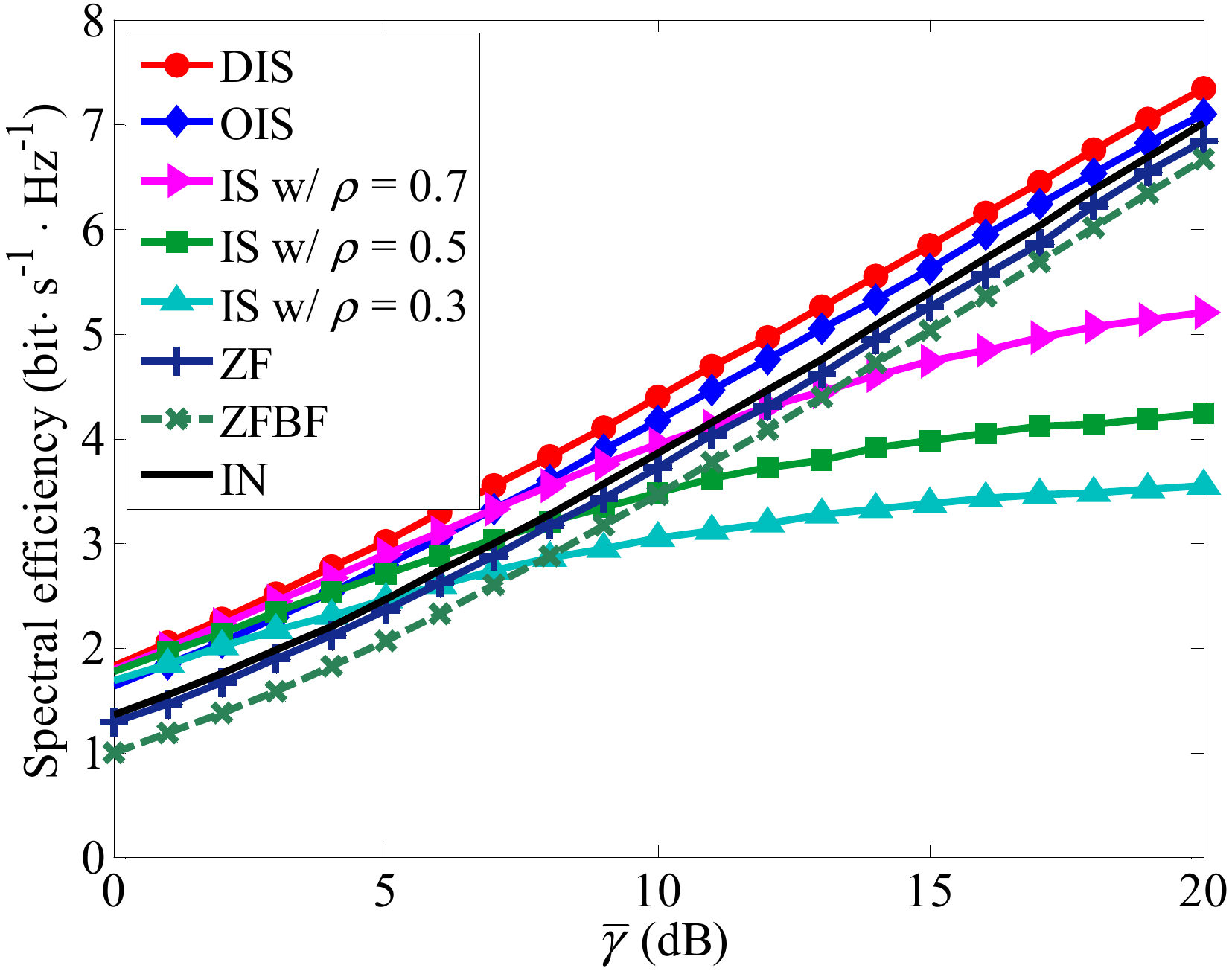}
\vspace{-16pt}
\caption{\fontsize{7.2pt}{\baselineskip}\selectfont{SE of PUE vs. $\bar{\gamma}$ with various IM schemes under $M=N=1$, and $\xi=1$.}}
%\caption{SE of PUE vs. $\bar{\gamma}$ with various IM schemes under $N_{T_{0}}=N_{T_{1}}=N_{R_{0}}=2$, $M=N=1$, and $\xi=1$.}
\label{fig: figure eleven}
\end{minipage}
\hfill
\begin{minipage}{0.66\columnwidth}
\centering
\includegraphics[width=1\columnwidth, height = 1.75in]{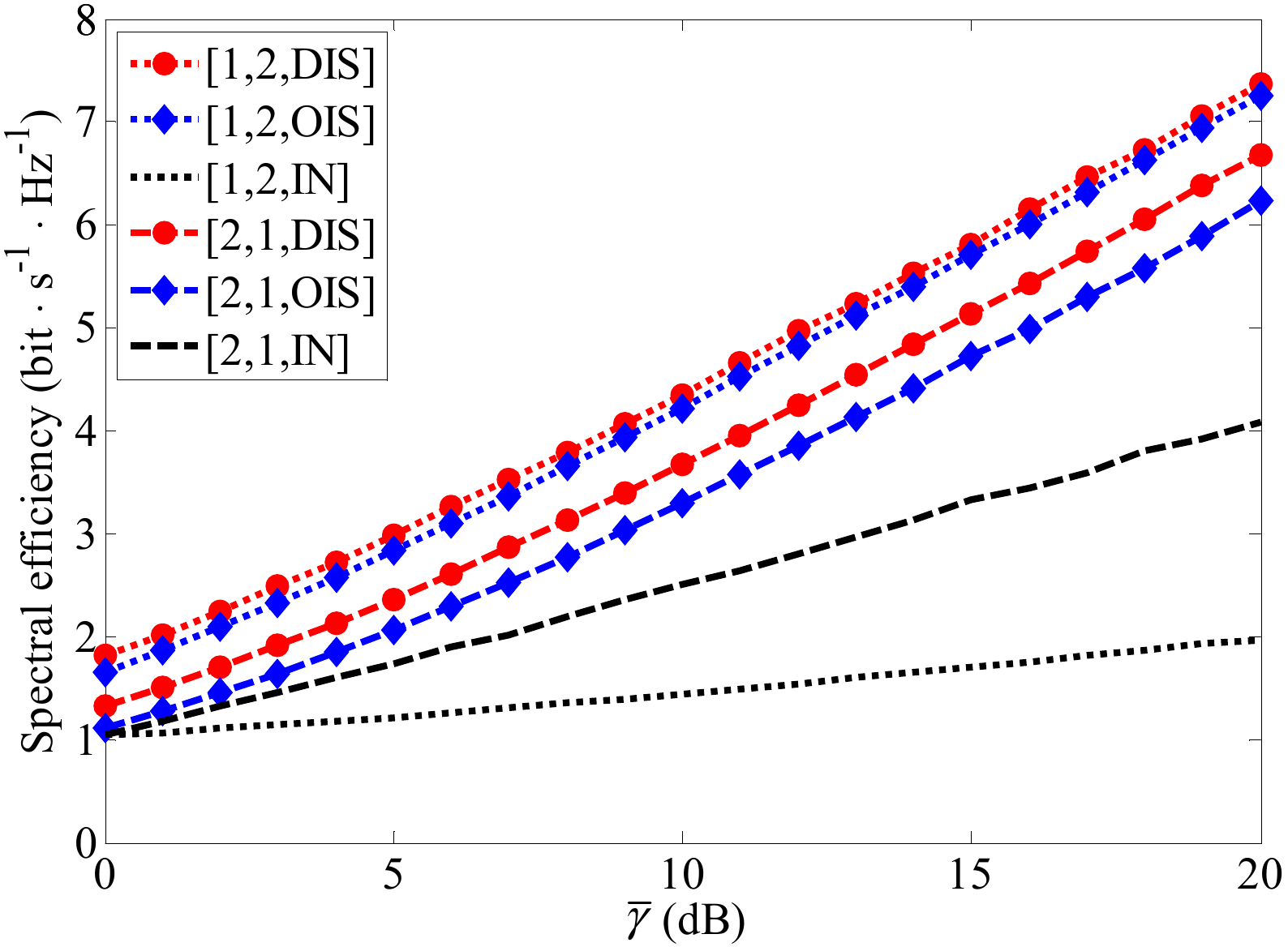}
\vspace{-16pt}
\caption{\fontsize{7.2pt}{\baselineskip}\selectfont{SE of PUE vs. $\bar{\gamma}$ with various IM schemes
under $\xi=1$ and different $M$ and $N$.}}
%\caption{SE of PUE vs. $\bar{\gamma}$ with various IM schemes under $N_{T_{0}}=N_{T_{1}}=N_{R_{0}}=2$, $\xi=1$ and different $M$ and $N$.}
\label{fig: figure twelve}
\end{minipage}
\vspace*{-0.1in}
\end{figure*}
%\begin{figure}[ht]
%%\vspace*{-0.1in}
%\graphicspath{/fig}
%\centering
%\includegraphics[width=0.33\textwidth]{fig/figure9b.pdf}
%%\vspace{-5pt}
%\caption{$Prob(P_{\mathcal{M}}>P_{0})$ vs. Power normalized by $P_{0}$ under $N_{T_{0}}=N_{T_{1}}=N_{R_{0}}=2$, $\xi=1$ and different $N$.}
%\label{fig: figure nine}
%%\vspace*{-0.15in}
%\end{figure}

Fig.~10 plots the distribution of $\bar{P}_{\mathcal{M}}$
for an arbitrary $\bar{\gamma}$ where
$\bar{P}_{\mathcal{M}}$ and $\bar{P}$ represents for
$P_\mathcal{M}$ and
the power value $P$ normalized by $P_{0}$.
Since $Prob(\bar{P}_{DIS}>\bar{P})$ varies with $\bar{\gamma}$ whereas
$Prob(\bar{P}_{IN}>\bar{P})$ and $Prob(\bar{P}_{OIS}>\bar{P})$ do not, for simplicity,
we only study IN and OIS.
As shown in the figure, $Prob(\bar{P}_{IN}>\bar{P})$ with $N=2$ is no less than that with $N=1$.
As for OIS, when $\bar{P}<0.5$, $Prob(\bar{P}_{OIS}>\bar{P})$ with $N=2$
is larger than that with $N=1$. However, as $\bar{P}$ grows larger than $0.5$,
$2$ interferences incur statistically less power overhead.
When $\bar{P}=1$, $Prob(\bar{P}_{\mathcal{M}}>\bar{P})$ with different schemes
shown in Fig.~10 is consistent with the results given in Fig.~9.
%\begin{figure}[ht]
%%\vspace*{-0.1in}
%\graphicspath{/fig}
%\centering
%\includegraphics[width=0.33\textwidth]{fig/figure7.pdf}
%%\vspace{-5pt}
%\caption{SE of PUE vs. $\bar{\gamma}$ with various IM schemes under $N_{T_{0}}=N_{T_{1}}=N_{R_{0}}=2$, $N=1$, and $\xi=1$.}
%\label{fig: figure twelve}
%%\vspace*{-0.15in}
%\end{figure}
%
%\begin{figure}[ht]
%%\vspace*{-0.1in}
%\graphicspath{/fig}
%\centering
%\includegraphics[width=0.33\textwidth]{fig/figure12.pdf}
%%\vspace{-5pt}
%\caption{SE of PUE vs. $\bar{\gamma}$ with various IM schemes under $N_{T_{0}}=N_{T_{1}}=N_{R_{0}}=2$, $\xi=1$, and different $N$.}
%\label{fig: figure twelve}
%%\vspace*{-0.15in}
%\end{figure}

Fig.~11 shows the PUE's average SE with different IM schemes.
Besides IN, OIS and DIS,
zero-forcing (ZF) reception and zero-forcing beamforming (ZFBF) are also simulated.
With ZF reception, a receive filter being orthogonal to the unintended signal is adopted
so as to nullify the interference at PUE, but an attenuation w.r.t. the desired signal results.
As for ZFBF, we let PBS adjust its beam so that the desired signal
is orthogonal to the interference at the intended receiver.
It should be noted that for either IN, IS with fixed $\rho$, OIS or DIS,
when the power overhead at PBS exceeds $P_{0}$,
i.e., the IM scheme is unavailable, we simply switch to ZF reception.
%The SE of DIS outperforms that of OIS, IN and IS with fixed $\rho\in \{0.3,0.5,0.7\}$.
As shown in the figure, DIS yields the best SE performance.
When $\bar{\gamma}$ is low, noise is the dominant factor affecting the PUE's SE.
Therefore, IS with fixed $\rho$ ($\rho<1$) yields similar %or even better
SE to OIS.
Moreover, although DIS can achieve the highest SE, its benefit is limited
in the low $\bar{\gamma}$ region.
As $\bar{\gamma}$ grows large, $\rho^{*}$ increases accordingly,
and hence IS with large $\rho$ exceeds that with small $\rho$ in SE.
In addition, OIS gradually outperforms those IS schemes with fixed $\rho$
as $\bar{\gamma}$ increases.
Moreover, by intellectually determining $\rho^{*}$,
the advantage of DIS becomes more pronounced with an increase of $\bar{\gamma}$.
Although IN yields more power overhead than IS with fixed $\rho$,
DIS and OIS, with the help of ZF, IN yields slightly higher SE than ZF reception.
As for ZFBF, more desired signal power loss results as compared to ZF,
thus yielding inferior SE performance.

Fig.~12 plots the PUE's average SE with various mechanisms under
different numbers of interferences and desired signals.
We use a general form $[M~N~\mathcal{M}]$ to denote the parameter settings,
where $M$ represents the number of desired signals,
$N$ is the number of interferences,
and $\mathcal{M}$ denotes the IM schemes.
When $M>1$, equal power allocation is adopted,
i.e., $P^{e}_{0,m}$ ($m=1,\cdots,M$) in Eq.~(21) is $P^{e}_{0}/M$.
As shown in the figure, OIS achieves the highest SE among
the three schemes, whereas IN yields the lowest SE.
Since $\rho^{*}$ approaches $1$ as $\bar{\gamma}$ increases,
with the same $M$ and $N$, DIS becomes OIS when $\bar{\gamma}$ grows too large.
Given fixed $\xi$, IN yields better SE when there are $2$ interfering signals
than the single interference case.
As for DIS and OIS, SE with $2$ interferences is lower than that with one disturbance.
This is consistent with the results shown in Figs.~9 and 10.

\section{Conclusion}

In this paper, we proposed a new interference management scheme,
called \textit{Dynamic Interference Steering} (DIS), for heterogeneous cellular networks.
By intelligently determining the strength of the steering signal,
the original interference is steered into an appropriate direction.
DIS can balance the transmit power used for generating the steering signal
and that for the desired signal's transmission.
Our in-depth simulation results show that the
proposed scheme makes better use of the transmit
power, and enhances users' spectral efficiency.

%In this paper, the solution of DIS for single interference has been given,
%yet for multi-interference situation,
%how to determine a set of optimal steering factors for
%all interfering components remains unsolved.
%Moreover, one should note that various IM schemes are
%implemented by different communication equipments
%at the cost of some performance loss, i.e., overhead.
%Then, how to comprehensively design IM mechanism from the system point of view
%so as to optimize the system's SE is also
%an important issue that warrants further investigation.
%In order to solve these problems,
%a universal model that can evaluate the benefits brought by,
%and the cost of an arbitrary IM scheme,
%as well as the conversion of different types of such benefits and cost,
%should be well developed.
%These are matters of our future inquiry.

% conference papers do not normally have an appendix

% use section* for acknowledgement
\section*{Acknowledgment}
This work was supported in part by NSFC (61173135,~U14\\05255);
the 111 Project (B08038); the Fundamental Research Funds for the Central Universities (JB171503).
It was also supported in part by the US National Science Foundation under Grant 1317411.

%\vspace{20pt}

% trigger a \newpage just before the given reference
% number - used to balance the columns on the last page
% adjust value as needed - may need to be readjusted if
% the document is modified later
%\IEEEtriggeratref{8}
% The "triggered" command can be changed if desired:
%\IEEEtriggercmd{\enlargethispage{-5in}}

% references section

% can use a bibliography generated by BibTeX as a .bbl file
% BibTeX documentation can be easily obtained at:
% http://www.ctan.org/tex-archive/biblio/bibtex/contrib/doc/
% The IEEEtran BibTeX style support page is at:
% http://www.michaelshell.org/tex/ieeetran/bibtex/
%\bibliographystyle{IEEEtran}
% argument is your BibTeX string definitions and bibliography database(s)
%\bibliography{IEEEabrv,../bib/paper}

\begin{thebibliography}{1}

\bibitem{CMYetis-1}
C.~M.~Yetis, T.~Gou, S.~A.~Jafar, et al., ``On feasibility of interference alignment in MIMO interference networks,''
IEEE Trans. Sig. Process., vol. 58, no. 9, pp. 4771-4782, 2010.

\bibitem{SJafar-2}
S.~Jafar and S.~Shamai, ``Degrees of freedom region of the mimo x channel,'' IEEE Trans. Inf. Theory, vol. 54, no. 1, pp. 151-170, 2008.

\bibitem{MMaddahAli-3}
M.~Maddah-Ali, A.~Motahari, and A.~Khandani, ``Communication over MIMO X channels: Interference alignment, decomposition, and performance analysis,''
IEEE Trans. Inf. Theory, vol. 54, no. 8, pp. 3457-3470, 2008.

\bibitem{VRCadambe-4}
V.~R.~Cadambe and S.~A.~Jafar, ``Interference alignment and degrees of freedom of the K-user interference channel,''
IEEE Trans. Inf. Theory, vol. 54, no. 8, pp. 3425-3441, 2008.

\bibitem{CSuh-5}
C.~Suh, M.~Ho, and D.~Tse, ``Downlink interference alignment,'' IEEE Trans. Commun., vol. 59, no. 9, pp. 2616-2626, 2011.

\bibitem{SMohajer-6}
S.~Mohajer, S.~N.~Diggavi, C.~Fragouli, and D.~N.~C.~Tse, ``Transmission techniques for relay-interference networks,''
in Proc. 46th Annu. Allerton Conf. Commun., Control, Comput., pp. 467-474, 2008.

\bibitem{SMohajer -7}
S.~Mohajer, S.~N.~Diggavi, and D.~N.~C.~Tse, ``Approximate capacity of a class of Gaussian relay-interference networks,''
in Proc. IEEE Int. Symp. Inf. Theory, vol. 57, no. 5, pp. 31-35, 2009.

\bibitem{TGou-8}
T.~Gou, S.~A.~Jafar, S.~W.~Jeon, and S.~Y.~Chung, ``Aligned interference neutralization and the degrees of freedom of the 2$\times$2$\times$2 interference channel,'' IEEE Trans. Inf. Theory, vol. 58, no. 7, pp. 4381-4395, 2012.

\bibitem{ZHo-9}
Z.~Ho and E.~A.~Jorswieck, ``Instantaneous relaying: optimal strategies and interference neutralization,'' IEEE Trans. Sig. Process., vol. 60, no. 12, pp. 6655-6668, 2012.

\bibitem{NLee-10}
N.~Lee and C.~Wang, ``Aligned interference neutralization and the degrees of freedom of the two-user wireless networks with an instantaneous relay,'' IEEE Trans. Commun., vol. 61, no. 9, pp. 3611-3619, 2013.

\bibitem{SBerger-11}
S.~Berger, T.~Unger, M.~Kuhn, A.~Klein, and A.~Wittneben, ``Recent advances in amplify-and-forward two-hop relaying,'' IEEE Commun. Mag., vol. 47, no. 7, pp. 50-56, 2009.

\bibitem{KGomadam-12}
K.~Gomadam and S.~Jafar, ``The effect of noise correlation in amplify-and-forward relay networks,'' IEEE Trans. Inf. Theory, vol. 55, no. 2, pp. 731-745, 2009.

\bibitem{ZliFGuo-13}
Z.~Li, F.~Guo, Kang G.~Shin, et al., ``Interference steering to manage interference,'' 2017. [Online].
Available: https://arxiv.org/abs/1712.07810. [Accessed Dec. 23, 2017].

\bibitem{TQSQuek-14}
T.~Q.~S.~Quek, G.~de la Roche, and M.~Kountouris, ``Small cell networks: deployment, PHY techniques, and resource management,'' Cambridge University Press, 2013.

\bibitem{WhitePaperCisco-15}
Cisco, ``Cisco Visual Networking Index: Global Mobile Data Traffic Forecast Update, 2015-2020,'' 2016.

\bibitem{FPantisano-16}
F.~Pantisano, M.~Bennis, W.~Saad, M.~Debbah, and M.~Latva-aho, ``Interference alignment for cooperative femtocell networks: A game-theoretic approach,''
IEEE Trans. Mobile Comput., vol. 12, no. 11, pp. 2233-2246, 2013.

\bibitem{3GPPTR-17}
3GPP TR 36.931 Release 13, ``LTE; Evolved Universal Terrestrial Radio Access (E-UTRA); Radio Frequency (RF) requirements for LTE Pico Node B,'' 2016.

\end{thebibliography}
%
% <OR> manually copy in the resultant .bbl file
% set second argument of \begin to the number of references
% (used to reserve space for the reference number labels box)

% that's all folks
\end{document}